\documentclass{birkjour}

\usepackage{amssymb,amsthm,slashed,mathabx,stackrel,ushort}

\usepackage[T1]{fontenc}
\usepackage[cp1250]{inputenc}

\usepackage{cite}
\usepackage{bm}
\usepackage{enumitem}

\usepackage[showonlyrefs]{mathtools}
\mathtoolsset{showonlyrefs=true}

\MakeRobust{\eqref}

\usepackage{bbm}

\usepackage{hyperref}
\usepackage[dvipsnames]{xcolor}

\hypersetup{colorlinks=false,
            hypertexnames=true,
            linkcolor=BlueViolet,
            linkbordercolor=LimeGreen,
            citecolor=Mahogany,
            citebordercolor=Goldenrod}

\newcommand{\Hc}{\mathcal{H}}
\newcommand{\mR}{\mathbb{R}}

\newcommand{\mN}{\mathbb{N}}
\newcommand{\lc}{C_+}
\newcommand{\lct}{C_+^t}
\newcommand{\wt}{\widetilde}
\newcommand{\al}{\alpha}
\newcommand{\vax}{\eta}
\newcommand{\sax}{\zeta}
\newcommand{\ep}{\epsilon}
\newcommand{\la}{\lambda}
\newcommand{\w}{\omega}
\newcommand{\W}{\Omega}
\newcommand{\dV}{\dot{V}}
\newcommand{\oV}{\bar{V}}
\newcommand{\doV}{\dot{\bar{V}}}
\newcommand{\dW}{\dot{W}}
\newcommand{\vep}{\varepsilon}
\newcommand{\dsp}{\displaystyle}
\newcommand{\ov}{\overline}
\newcommand{\p}{\partial}
\newcommand{\con}{\mathrm{const}}
\newcommand{\rg}{\mathrm{rg}}
\newcommand{\ar}{\mathrm{ar}}
\newcommand{\qu}{\mathrm{q}}
\newcommand{\ax}{\widearrow{ax}}
\newcommand{\yx}{\widearrow{yx}}
\newcommand{\dm}{\mathrm{d}}
\newcommand{\x}{\mathbf{x}}
\newcommand{\slf}{\mathrm{sl}}

\DeclareMathOperator{\id}{\mathbbm{1}}
\DeclareMathOperator{\sgn}{sgn}
\DeclareMathOperator{\supp}{supp}

\setlist[itemize]{leftmargin=2em}

\begin{document}

\title{Almost radial gauge}

\author{Andrzej Herdegen}

\address{Institute of Theoretical Physics\\
    Jagiellonian University\\
    ul.\,S.\,{\L}ojasiewicza 11\\
    30-348 Krak\'{o}w\\
    Poland}

\email{herdegen@th.if.uj.edu.pl} {\it}\date{}



\date{}

\begin{abstract}

An almost radial gauge $A^\ar$ of the electromagnetic potential is
constructed for which $x\cdot A^\ar(x)$ vanishes arbitrarily fast in
timelike directions. This potential is in the class introduced by Dirac
with the purpose of forming gauge-invariant quantities in quantum
electrodynamics. In the quantum case, the construction of smeared operators
$A^\ar(K)$ is enabled by a~natural extension of the free electromagnetic
field algebra introduced earlier (represented in a Hilbert space). The
space of possible smearing functions $K$ includes vector fields with the
asymptotic spacetime behavior typical for scattered currents (the
conservation condition in the whole spacetime need not be assumed). This
construction is motivated by a possible application to the infrared problem
in QED.

\end{abstract}

\maketitle

\section{Introduction}\label{int}

The long-range nature of interaction entails the so called infrared problems
in electrodynamics, both classical and quantum \cite{jau76}. Some of the
quantum problems in this area are specific for this realm, but other reveal
themselves already on the classical level. The identification of asymptotic
charged fields and the scattering theory belong to the latter class.

The present author has long followed the idea that an appropriate choice of
the electromagnetic gauge may relieve the scattering infrared problems. After
preliminary results \cite{her95}, it was shown recently that indeed the
classical asymptotic problem for scattering of the Dirac field in
electromagnetic time-dependent field, typical for the complete theory,
disappears in certain gauges. The main constituent feature of this class is
sufficiently fast vanishing of the product $x\cdot A(x)$ for $x$ tending to timelike infinity (with an arbitrarily fixed origin $x=0$ in Minkowski space) \cite{her21}. The choice of a specific potential in this class has a large freedom in this classical context. However, we plan to extend our analysis to quantum electrodynamics, and in that case a judicious choice of our gauge is of fundamental importance.

The first gauge which probably comes to mind is the radial gauge
\mbox{$x\cdot A^\rg(x)=0$}, going also in literature by the names
Fock-Schwinger or relativistic Poincar\'e gauge (\!\cite{foc37,sch51},
see also \cite{jac02}). In classical electrodynamics this gauge is well
defined and may be obtained by integration of the electromagnetic field
tensor. In quantum theory the latter property would be an advantage, as no
indefinite metric space would be needed. However, we shall indicate in the
following that the quantum version of this gauge is singular, which explains
the need for \emph{a posteriori} regularization in attempts to construct
perturbative calculus with the use of this gauge \cite{leu96}.

It is the purpose of the present article to construct an `almost radial
gauge,' which may also be obtained from the electromagnetic field tensor
alone, and for which $x\cdot A^\ar(x)$ vanishes fast in timelike directions,
but which at the same time admits quantization. Almost radial gauge turns out
to be in the Dirac family of `gauge-invariant gauges', which in the standard
formulation of QED may be constructed only after initial regularization and
the use of the indefinite Gupta-Bleuler metric (see the discussion of this
problem in \cite{sta00}, p.~191). We choose a different approach. For the
construction of the quantum field we use a natural extension of the algebra
of the free field proposed earlier \cite{her98}, which enables representation
of infrared-singular fields.  This extension admits the construction of the
almost radial gauge, and allows moreover its well-defined smearing with
vector functions typically appearing in scattering.

As we shall see, the construction of the almost radial gauge involves
integration of the electromagnetic field along radial $\mR$-axes. This bears
some similarity to the integration of the field along spacelike semi-axes,
which is the principle of the construction of string-localized potentials
introduced by Mund, Schroer and Yngvason \cite{msy06}. We shall discuss
similarities and differences of the two constructions.

The plan of the article is as follows. In Section \ref{pre} we summarize our
notation and mathematical tools to be used in the article. Section \ref{fqf}
contains a brief summary of the quantization of the standard free field
algebra, and its extension mentioned above. The classical radial gauge, and
its singularity in quantum case, are discussed in Section \ref{rg}. Section
\ref{arg} contains our main results: construction and evaluation of
properties of the almost radial gauge, both classical and quantum, along the
lines announced above. In Section \ref{dis} we compare our construction with
the string-localized potentials. Section \ref{outlook} contains some final remarks. Proofs of some technical points are shifted to Appendix.

\section{Preliminaries}\label{pre}
We list here our notation, conventions and preliminary formulas. More
extensive summary of these questions may be found in \cite{her17}.

We consider the flat spacetime with a fixed origin, thus described by the
Min\-kow\-ski vector space $M$ with the scalar product $x\cdot y$ with
signature $(+,-,-,-)$. We choose physical units for which $\hbar$, $c$ and an
arbitrarily chosen length scale are all equal to $1$. Then, in particular,
$x\in M$ may be treated as dimensionless. We fix a unit timelike,
future-pointing vector $t$ and for a~Min\-kow\-ski vector $x$ we write
$x^0=x\cdot t$ and $\x=x-(x\cdot t)t$. The $3$-dimensional norm of $\x$ is
denoted $|\x|$, and we also write $|x|=(|x^0|^2+|\x|^2)^{\frac{1}{2}}$. The
tensor indices, which are often suppressed, are denoted by $a$, $b$ etc. The
Minkowski volume measure element is denoted $\dm x$. Our results do not
depend on the choice of vector $t$.

\subsection{Homogeneous functions on the future light cone}\label{hf}

We shall write $l$ for any future-pointing lightlike (nonzero) vector and
denote
\begin{equation}\label{pre-lc}
 \lc=\{\,l\in M\mid l\cdot l=0,\ l^0>0\,\}\,,\quad \lct=\{\,l\in\lc\mid l^0=1\,\}\,.
\end{equation}
For $C^1$-functions $f(l)$ on $C_+$, the intrinsic differentiation operators
\begin{equation}
 L_{ab}=l_a\frac{\p}{\p l^b}-l_b\frac{\p}{\p l^a}
\end{equation}
are the generators of the Lorentz transformations $f(l)\mapsto
f_\Lambda(l)=f(\Lambda^{-1}l)$, where $\Lambda$ is a Lorentz transformation
of the Minkowski space. The derivative $\p f/\p l^a$ can only be given a unique meaning if one determines a particular $C^1$-extension of $f$ to a
neighborhood of $C_+$. However, any two such extensions differ by a term
$l^2g(l)$, with a $C^1$-function $g(l)$ in a neighborhood of $\lc$. Thus, on
$\lc$ the derivatives $\p f(l)$ obtained from these two extensions differ by
$2lg(l)$. Therefore, if we define
\begin{equation}\label{pre-defpart}
 \p_a f(l)=(l^0)^{-1}L_{0a} f(l)
 \end{equation}
in any chosen reference frame, then $\p_a f(l)$ is uniquely defined as an
equivalence class with respect to the addition of terms proportional to
$l_a$. If the definition of $f(l)$ is the restriction of a function naturally
defined in some neighborhood of $\lc$, then this function may also serve to
define $\p_af(l)$.

For any measurable function $W(l)$ on $\lc$, homogeneous of degree $-2$, the
integral over the set of null directions defined by
\begin{equation}\label{pre-d2l}
 \int W(l)\,\dm^2l=\int_{\lct} W(l)\,\dm\W_t(l)\,,
\end{equation}
where $\dm\W_t(l)$ is the angle measure on the unit sphere, does not depend
on the choice of the vector $t$. If $W$ is $C^1$, it follows that
\begin{equation}\label{pre-iL}
 \int L_{ab}W(l)\,\dm^2l=0\,.
\end{equation}
Another closely related identity is the following. Let $V^a(l)$ be a
$C^1$-vector function on $\lc$, such that $l\cdot V(l)=0$. Suppose $V(l)$ is
extended to a~neighborhood of the light cone with the preservation of these
properties. Then in each Minkowski frame one has
\begin{equation}
 L_{0a}\bigg(\frac{V^a}{l^0}\bigg)=\p\cdot V(l)\,,
\end{equation}
so the rhs of this equation is both frame- and extension (preserving
ortho\-go\-na\-li\-ty)-independent. Moreover, if in addition $V(l)$ is
homogeneous of degree~$-1$, then
\begin{equation}\label{pre-idV}
 \int \p\cdot V(l)\,\dm^2l=0\,.
\end{equation}
If $f(l)$ is a scalar $C^1$-function, homogeneous of degree $0$, then the
last identity and the earlier remarks on $\p_af(l)$ imply the `integration by
parts' identity
\begin{equation}\label{pre-parts}
 \int f(l)\,\p\cdot V(l)\,\dm^2l=-\int V(l)\cdot \p f(l)\,\dm^2l\,.
\end{equation}

We end these remarks with the following definition. Let $F(l)$ be any
continuous function, homogeneous of degree $-1$. Then it is easily calculated
that
\begin{equation}
 \lim_{\ep \searrow 0}\,\frac{1}{2\pi} \int \delta(l\cdot l'-\ep l^0{l'}^0)F(l')\,\dm^2l'=F(l)\,,
\end{equation}
independently of the frame in which the coordinates $l^0{l'}^0$ are taken,
where $\delta(.)$ is the Dirac delta distribution. Therefore, we can
interpret the lhs as the action of a distribution $\delta(l,l')$, homogeneous
of degree $-1$ in each of the arguments:
\begin{equation}
 \int \delta(l,l')F(l')\,\dm^2l'=F(l)\,.
\end{equation}

\subsection{Fourier transforms}
We use the following conventions for a function $f$ on $M$ and a function $g$ on $\mR\times \lc$:
\begin{equation}\label{pre-Fou}
\begin{aligned}
 \hat{f}(p)&=\frac{1}{2\pi}\int_M f(x) e^{ip\cdot x}\dm x\,,\\
 \wt{g}(\w,l)&=\frac{1}{2\pi}\int_\mR g(s,l)e^{i\w s}\dm s\,.
\end{aligned}
\end{equation}

\subsection{Wave equation}\label{pre-we}
Solutions of the homogeneous wave equation may be represented by
\begin{equation}\label{pre-Adl}
 A(x)=-\frac{1}{2\pi}\int \dV(x\cdot l,l)\,\dm^2 l\,,
\end{equation}
where\footnote{In what follows, the overdot will always denote the
differentiation with respect to the real parameter $s$; for a choice of
$V(s,l)$ see below.} $\dV(s,l)$ (of any algebraic type) is a function
homogeneous of degree $-2$. This representation is related to the Fourier
representation
\begin{equation}
 A(x)=\frac{1}{\pi}\int_M e^{-ix\cdot k}a(k)\sgn(k^0)\delta(k^2)\,\dm k
\end{equation}
by
\begin{equation}\label{pre-Va}
 \wt{\dV}(\w,l)=-\w a(\w l)\,.
\end{equation}
In physical contexts of interest the function $\dV(s,l)$ is of class $C^N$,
for some $N\in\{0,1,\ldots\}\cup\{\infty\}$, and for $l$ scaled\footnote{This
scaling is to be understood in all further estimates, to appear below,
containing $s$ or~$l$.} to $l^0=1$ and some $\vep>0$ the following bounds
hold:
\begin{equation}\label{pre-Vbound}
 |L_{a_1b_1}\ldots L_{a_kb_k}\dV(s,l)|\leq\frac{\con}{(1+|s|)^{1+\vep}}\,,\qquad
 k=0,1,\ldots,N\,.
\end{equation}
With this condition, $A(x)$ has null asymptotes given by
\begin{equation}\label{pre-as}
 \lim_{R\to\infty}RA(x\pm Rl)=\pm \big[V(x\cdot l,l)-V(\pm\infty,l)\big]\,,
\end{equation}
where $V(s,l)$ is any primitive function of $\dV(s,l)=\p V(s,l)/\p s$,
determined up to an additive $s$-independent term. Independent of this choice
is the difference
\begin{equation}\label{pre-DV}
 \Delta V(l)=V(+\infty,l)-V(-\infty,l)=\int_\mR\dV(s,l)\,\dm s\,.
\end{equation}
In what follows two spaces of functions $V(s,l)$ will appear naturally: one
characterized by $V(+\infty,l)=0$, and another by $V(-\infty,l)=0$. In addition, we shall also find convenient to define another
choice, denoted by
\begin{equation}\label{pre-MV}
 \oV(s,l)=\tfrac{1}{2}\int_\mR \sgn(s-\tau)\dV(\tau,l)\,\dm \tau\,,
\end{equation}
for which
\begin{equation}
 \oV(\pm\infty,l)=\pm\tfrac{1}{2}\Delta V(l)\,.
\end{equation}

Beside the existence of null asymptotes, condition \eqref{pre-Vbound} entails
(with the use of estimate \eqref{apes-es}) the following decay of $A(x)$ in
spacetime:
\begin{equation}\label{pre-decA}
\begin{aligned}
 |A(x)|&\leq\con\int\frac{\dm\Omega_t(l)}{(1+|x\cdot l|)^{1+\vep}}
 =\int_{-1}^1\frac{\con\,\dm u}{(1+||x^0|+|\x|u|)^{1+\vep}}\\[1ex]
 &\leq \frac{\con}{1+|x^0|+|\x|}
 \bigg\{\theta(-x^2)+\frac{\theta(x^2)}{(1+|x^0|-|\x|)^\vep}\bigg\}\,.
\end{aligned}
\end{equation}

We end this section by mentioning that the standard Pauli-Jordan function, a
particular solution of the wave equation, has the representation
\begin{equation}\label{pre-PJ}
 D(x)=\frac{1}{2\pi}\sgn(x^0)\delta(x^2)=-\frac{1}{8\pi^2}\int \delta'(x\cdot l)\,\dm^2l\,.
\end{equation}
This may be shown by writing the rhs as
\begin{equation}
 -\frac{1}{8\pi^2}\frac{\p}{\p x^0}\int \frac{\delta(x\cdot l)}{l^0}\,\dm^2l
 =-\frac{1}{4\pi}\frac{\p}{\p x^0}\int_{-1}^1 \delta(x^0+|\mathbf{x}|u)\dm u
 =-\frac{1}{4\pi}\frac{\p}{\p x^0}\frac{\theta(-x^2)}{|\mathbf{x}|}\,,
\end{equation}
which gives the standard form.

\subsection{Free electromagnetic field}\label{free}

A Lorenz potential $A$ of a free electromagnetic field $F$ has the
re\-pre\-sen\-ta\-tion~\eqref{pre-Adl}, where $\dV(s,l)$ is a vector function
satisfying the condition
\begin{equation}\label{pre-ldotV}
l\cdot\dV(s,l)=0\,.
\end{equation}
We assume that the condition \eqref{pre-Vbound} with $N\geq1$ is satisfied,
so $\Delta V(l)$ exists, and satisfies $l\cdot \Delta V(l)=0$. Moreover, we
demand in addition that
\begin{equation}\label{pre-LV}
 L_{[ab}\Delta V_{c]}(l)=0\,,
\end{equation}
which under the orthogonality condition is equivalent to the existence of a
homogeneous of degree $0$ function $\Phi(l)$, such that
\begin{equation}\label{pre-VdFi}
 \Delta V_a(l)=-\p_a\Phi(l)\,,
\end{equation}
equality in the sense described in Section \ref{hf}. Then $A$ is of the type
of radiation potentials of fields created in scattering processes. The
corresponding electromagnetic field is then given by
\begin{equation}\label{pre-F}
 F_{ab}(x)=-\frac{1}{2\pi}\int\big[ l_a\ddot{V}_b(x\cdot l,l)-l_b\ddot{V}_a(x\cdot l,l)\big]\,\dm^2l\,.
\end{equation}
The gauge freedom within the space of Lorenz gauges is represented in terms
of $\dV(s,l)$ by the transformations
\begin{equation}\label{pre-gauge}
 \dV(s,l)\mapsto \dV(s,l)+l\dot{\alpha}(s,l)\,,
\end{equation}
 with $\alpha(s,l)$ homogeneous
of degree $-2$. The spacelike asymptotic behavior of $A$ (and $F$) depends
on $\Delta V(l)$ and is given, independently of $x$,  by
\begin{equation}\label{pre-tail}
 \lim_{R\to\infty} RA(x+Ry)= -\frac{1}{2\pi}\int \Delta V(l)\,\delta(y\cdot l)\,\dm^2l\,,\qquad
 y^2<0\,.
\end{equation}
Fields with $\Delta V(l)\neq 0$ are usually termed as infrared singular;
their spacelike tail falls off as that of the Coulomb field of the electric
charge.

For future use we note the following representation
\begin{equation}\label{pre-xA}
 x\cdot A(x)=\frac{1}{2\pi}\int (\p\cdot V)(x\cdot l,l)\,\dm^2l\,;
\end{equation}
here and in what follows we denote
\begin{equation}
 (\p_a V)(x\cdot l,l)=\frac{\p}{\p l^a} V(s,l)|_{s=x\cdot l}
\end{equation}
(and similarly for $L_{ab}$ and other functions in place of $V$).
Representation \eqref{pre-xA} follows immediately from the relation
\begin{equation}
 \frac{\p}{\p l}\cdot V(x\cdot l,l)=x\cdot\dV(x\cdot l,l)+(\p\cdot V)(x\cdot l,l)
\end{equation}
by the use of identity \eqref{pre-idV}.

Finally, it is important for what comes to relate representation
\eqref{pre-Adl} to the construction of the radiation potential of a conserved
vector current $J(x)$,
\begin{equation}
 \p\cdot J(x)=0\,.
\end{equation}
Let $J(x)$ be such current, for which $\dV(s,l)$, with
\begin{equation}\label{pre-VJ}
 V(s,l)=\int J(x)\delta(s-x\cdot l)\,\dm x\,,
\end{equation}
satisfies conditions \eqref{pre-Vbound}. Then the radiation potential of this
current has two equivalent forms
\begin{equation}\label{pre-radA}
 A[J](x)=4\pi\int_M D(x-y)J(y)\,\dm y=-\frac{1}{2\pi}\int \dV(x\cdot l,l)\,\dm^2l\,,
\end{equation}
which is shown with the use of representation \eqref{pre-PJ}. We note that
orthogonality \eqref{pre-ldotV} is then automatic, but in general
\begin{equation}\label{pre-Q}
 l\cdot V(s,l)=Q\,,
\end{equation}
$Q$ being the charge of the current.

\section{Free quantum field}\label{fqf}

Quantization of the local free electromagnetic field may be described in
terms of the  algebra of electromagnetic potential elements $A(J)$ satisfying
the relations
\begin{equation}\label{fqf-comAJ}
 [A(J_1),A(J_2)]=i\,4\pi\int J_1(x)D(x-y)J_2(y)\,\dm x\,\dm y=i\{J_1,J_2\}\,,
\end{equation}
where $J_i$ are smooth, compactly supported conserved test currents (as
discussed in \cite{her98} and \cite{her08}). After exponentiation to the Weyl
form, this defines a~net of $C^*$-algebras satisfying Einstein causality in
the form of commutation for spacelike separated supports of
$J_i$.\footnote{More recently, description in terms of `intrinsic potential'
$A(J)$ has been used for an axio\-matic construction of a general local
electromagnetic algebra in \cite{buch15}. In that case, according to this
analysis, the commutators for spacelike separated, but topologically
nontrivial regions need not vanish, but are in the center of the algebra.}
This formulation is close to the discussion given in \cite{roe70}. In more
standard formulation one treats the free electromagnetic field as an operator-valued distribution on the space of antisymmetric, smooth, compactly
supported functions $\varphi^{ab}(x)$. Then the elements
\begin{equation}\label{fqf-Ffi}
 F(\varphi)=A(J)\,,\quad \text{where}\quad J^a=2\p_b\varphi^{ab}\,,
\end{equation}
generate the standard net of electromagnetic field, where now $F(\varphi)$ is
considered as an element of the local algebra of any open set containing
$\supp\varphi$. However, as each conserved, compactly supported current $J^a$
has a representation \eqref{fqf-Ffi} with a compactly
supported\footnote{\label{poinc}This more specific Poincar\'e lemma for
compact supports follows from its standard form and the properties of
cohomology of spheres, see e.g.\ \cite{lee13}, Lemma 17.27. Also, the proof
in that reference easily implies that for $\supp J$ contained in an open,
star-shaped set, there exists $\varphi$ with support contained in the same
set. See also \cite{buch15}.} $\varphi^{ab}$, the nets have the same
elements, with some differences in the assignment to spacetime sets.
Moreover, two important properties bind the two nets:
\begin{itemize}
\item[(i)] the nets are local with respect to each other, i.e.,
\begin{equation}\label{fqf-FAloc}
 [F(\varphi),A(J)]=0
\end{equation}
for spacelike separated supports of $\varphi$ and $J$;
\item[(ii)] for each open star-shaped set the two local algebras assigned
    to this set are equal.
\end{itemize}
The first property  is obvious from \eqref{fqf-comAJ} and \eqref{fqf-Ffi},
and for the second see footnote \ref{poinc}.

The symplectic form on the rhs of \eqref{fqf-comAJ} may be expressed in terms
of functions $V(s,l)$ defined by \eqref{pre-VJ}. Substituting in this form
representation \eqref{pre-PJ}, one obtains\footnote{By a slight abuse of
notation, we use here the same symbol as in \eqref{fqf-comAJ}.}
\begin{equation}\label{fqf-sympl}
\begin{aligned}
 \{V_1, V_2\}
 &=\frac{1}{4\pi}\int\big[\dV_1(s,l)\cdot V_2(s,l)-V_1(s,l)\cdot \dV_2(s,l)\big]\,\dm s\,\dm^2l\\
 &=i \int \ov{\wt{\dV}_1(\w,l)}\cdot \wt{\dV}_2(\w,l)\,\frac{\dm\w}{\w}\dm^2l\,,
\end{aligned}
\end{equation}
where for the Fourier transformed version we used \eqref{pre-Fou}. If one
writes symbolically $A(J_i)=\int_M A(x)J_i(x)\,\dm x$, with $A(x)$
represented by \eqref{pre-Adl} with quantum function $V^\qu(s,l)$, then
\begin{equation}\label{fqf-VVq}
 A(J_i)=\{V_i, V^\qu\}\,.
\end{equation}
The quantization condition \eqref{fqf-comAJ} takes now the form
\begin{equation}\label{fqf-comWW}
 \big[\{V_1,V^\qu\},\{V_2,V^\qu\}\big]=i\{V_1, V_2\}\,.
\end{equation}
This condition is now extended to the space of all smooth, compactly
supported functions $V(s,l)$, satisfying orthogonality\footnote{This
formulation of the quantization is in the spirit of `asymptotic quantization'
discussed by Bramson \cite{bra77} and later by Ashtekar \cite{ash86}).}\
\begin{equation}\label{fqf-ortho}
 l\cdot V(s,l)=0\,,
\end{equation}
which for $V(s,l)$ obtained from $J$ reflects the fact that the current is
charge-less. The form \eqref{fqf-sympl} is nondegenerate on the space of
equivalence classes of such functions with respect to gauge transformations
\eqref{pre-gauge}.  This may be also interpreted as follows: the symplectic
structure is defined and nondegenerate on the space of infrared regular
classical fields \eqref{pre-F}. This leads to the unique Weyl algebra
satisfying the corresponding exponentiated condition on this space (see
\cite{bra96}, Section 5.2.2.2). Consider the scalar product on the space of
equivalence classes
\begin{equation}\label{fqf-scpro}
 (\wt{\dV}_1,\wt{\dV}_2)
 =\int\int_{\mR_+} \Big[-\ov{\wt{\dV}_1(\w,l)}\cdot \wt{\dV}_2(\w,l)\Big]\,\frac{\dm\w}{\w}\dm^2l\,,
\end{equation}
and the Hilbert space $\Hc$ obtained by completion. The standard construction
leads to the representation of the algebra on the Fock space based on $\Hc$.
When restricted to the exponentiated elements $F(\varphi)$, this is the
standard irreducible vacuum representation of the local net.

In interacting theory one needs to form elements $A(K)$, where the vector
function $K$ need not be conserved, and secondly, need not be compactly
supported. The first problem (non-conservation of $K$) is dealt by choosing
$A$ in some special gauge, and in local formulation of QED this leads to the
Gupta-Bleuler indefinite metric. The second problem (non-compactness) is
avoided in usual formulations by the introduction of a switching function
$g(x)$ and the subsequent performance of the `adiabatic limit' $g\to 1$. This
programme of the solution of infrared problems is far from
complete.\footnote{But see a recent contribution to this programme
\cite{duc19}.} Here we want to propose a construction avoiding both problems,
with the use of a natural extension of the above algebra to the free
infrared singular fields, put forward in~\cite{her98}. We briefly sketch the
construction of this extension.

The symplectic structure \eqref{fqf-sympl} has a natural extension to a wider
space. It is easy to see that the form remains well defined (absolutely
integrable) and gauge invariant for general functions $V_i$ subject to
conditions \eqref{pre-Vbound} with $N=\infty$, \eqref{pre-VdFi} and
\eqref{fqf-ortho}. However, for this structure to be again consistently
formulated on the space of all classical fields \eqref{pre-F} (now also those
infrared-singular when $\Delta V_i(l)\neq0$), we have to make the choice of
$V(s,l)$ for given $\dV(s,l)$ unique. We make this choice guided by the
following physical motivation (as discussed in \cite{her08}). Free fields
appear in realistic physical theories as incoming or outgoing fields in
scattering setting. We want to admit the possibility of testing these fields
with currents infinitely extended in timelike directions.\footnote{Extensive
discussion of such test currents may be found in \cite{her08}; see also what
follows.} Formula \eqref{pre-VJ} tells us that for $V(+\infty,l)$
($V(-\infty,l)$) to be different from zero, the test current $J$ must have a
sufficiently slow (in fact, of $|x|^{-3}$ type) decay in the future (past),
respectively. Therefore, it seems reasonable to adopt the rule that
\begin{equation}\label{fqf-Vinout}
\begin{aligned}
 &V(+\infty,l)=0 & &\text{for `in' case}\,,\\
 &V(-\infty,l)=0 & &\text{for `out' case}\,,
\end{aligned}
\end{equation}
consistently with the spacetime regions where the fields are tested: `in' in
the past, and `out' in the future; the same rule is adopted for the quantum
variables $V^\qu(s,l)$.  Let us mention that all such extended currents are
charge-free (see \eqref{pre-Q}), and that the symplectic form $\{V_1,V_2\}$
for such currents may be represented in the form
\begin{equation}\label{fqf-JJcom}
 \{J_1,J_2\}=\tfrac{1}{2}\int_M \big(A[J_1]\cdot J_2-A[J_2]\cdot J_1\big)(x)\,\dm x\,,
\end{equation}
where $A[J_i](x)$ are radiation fields \eqref{pre-radA}. For compactly
supported $J_i$ this agrees with the definition in \eqref{fqf-comAJ};
however, the latter is not absolutely integrable in general.

Relations \eqref{fqf-comWW}, considered on the space of functions $V(s,l)$
satisfying \eqref{pre-Vbound} with $N=\infty$, \eqref{pre-VdFi},
\eqref{fqf-ortho} and \eqref{fqf-Vinout}, define (after exponentiation) a
$C^*$-algebra, an extension of the standard local algebra. Algebras for `in'
and `out' cases are canonically isomorphic. For $J_i$ with supports spacelike
separated the usual commutation rule is fulfilled. Also, the form
\eqref{fqf-JJcom} shows that the timelike commutativity characteristic for
massless fields extends as well: if $\supp J_1$ is timelike separated from
$\supp J_2$, then the two fields commute, irrespective of the size of the
supports.

The Poincar\'e group transformations $(z,\Lambda)$ are represented by
the automorphisms $\alpha_{z,\Lambda}$ acting on the algebra by
\begin{equation}\label{fqf-poinc}
 \alpha_{z,\Lambda}\big[A(J)\big]=A\big(T_{z,\Lambda}J\big)\,,
\end{equation}
where $T_{z,\Lambda}$ is the natural representation of the Poincar\'e group
on tensor functions. Expressed in terms of notation \eqref{fqf-VVq}, this
reads
\begin{equation}\label{fqf-poinVV}
 \alpha_{z,\Lambda}[\{V,V^\qu\}]=\{T_{z,\Lambda}V,V^\qu\}\,,
\end{equation}
where
\begin{equation}\label{fqf-poinV}
 [T_{z,\Lambda}V]_a(s,l)=\Lambda_a{}^b V_b(s-z\cdot l,\Lambda^{-1}l)\,.
\end{equation}

Translationally covariant, positive energy, irreducible representations of
the extended algebra have been constructed in \cite{her98} and~\cite{her08}.
Also, there exist among them those for which rotations are implemented,
although not Lorentz boosts. In these representations there are vector states
with arbitrarily small energy content, but no vacuum vector state exists.

The integral in the Fourier-transformed version of the symplectic form
\eqref{fqf-sympl} is now in the principal value sense. Note also that
$\wt{\dV}(0,l)=(2\pi)^{-1}\Delta V(l)$. Therefore, elements $\{V_1,V^\qu\}$
with $\Delta V_1(l)\neq0$, describing infrared-singular fields, have no
representation in the vacuum representation of the  local algebra, which is
based on the scalar product \eqref{fqf-scpro}. We shall see in
Section~\ref{arg} below, that our larger arena enables the construction of
elements $A^\ar(K)$ for appropriately extended functions $K$ and almost
radial gauge $A^\ar$ to be defined there.

For later comparison, we write the symbolic distributional formula equivalent
to \eqref{fqf-comWW} (for test functions $V_i$ vanishing at infinity):
\begin{equation}\label{fqf-comV}
 \big[\dV^\qu_a(s,l),\dV^\qu_b(s',l')\big]
 =-i\,g_{ab}\,\frac{2\pi}{l^0{l'}^0}\,\delta(l,l')\,\delta'\Big(\frac{s}{l^0}-\frac{s'}{{l'}^0}\Big)\,.
\end{equation}
The rhs has the correct homogeneity and is independent of the choice of the
vector $t$. The metric tensor on the rhs could suggest indefinite metric, but
this is not the case: this relation is smeared with (equivalence classes with
respect to \eqref{pre-gauge} of) functions $V_i(s,l)$ orthogonal to $l$. In
local formulation of interacting theory one needs to smear $A(x)$ with
non-conserved test vector fields, and then the usual Gupta-Bleuler
formulation is needed.

Finally, let us note for future use that if both $V_i(s,l)$ satisfy one and
the same of the conditions \eqref{fqf-Vinout}, then
\begin{equation}\label{fqf-VbarV}
 \{V_1,V_2\}=\{\oV_1,\oV_2\}\,,
\end{equation}
where $\oV_i(s,l)$ are formed by the convention \eqref{pre-MV}.

\section{Radial gauge}\label{rg}

As announced in Introduction, we are now looking for a suitable quantum
gauge, in which $x\cdot A(x)$ vanishes (sufficiently fast) in timelike
directions. We start by testing the radial gauge for that purpose, for a
classical field first.

If $F_{ab}(x)=\p_aA_b(x)-\p_bA_a(x)$, with $A$ a $C^1$-function, then without
any further conditions the following identity holds:
\begin{equation}
 x^aF_{ab}(x)=(x\cdot\p+1)A_b(x)-\p_b(x\cdot A(x))\,,
\end{equation}
so for $\la\in\mR$ we have
\begin{equation}\label{rg-ident}
 \la x^a F_{ab}(\la x)=\frac{\p}{\p\la}\big[\la A_b(\la x)\big]-\p_b(x\cdot A(\la x))\,.
\end{equation}
Integrating with respect to $\la$ on $[0,1]$ one finds
\begin{align}
 \int_0^1 \la x^a F_{ab}(\la x)\dm \la
 &= A_b(x)-\p_bS^\rg(x)\equiv A^\rg_b(x)\,,\label{rg-defrad}\\
 S^\rg(x)&=\int_0^1 x\cdot A(\la x)\dm\la\,,
\end{align}
which for any electromagnetic field $F_{ab}$ defines the radial gauge with
the property
\begin{equation}\label{rg-FS}
 x\cdot A^\rg(x)=0\,.
\end{equation}
Potential \eqref{rg-defrad} is the unique regular solution of the condition
\eqref{rg-FS}. Indeed, any addition to $A_a^\rg(x)$ of a gauge term
$\p_aG(x)$ respecting \eqref{rg-FS} must result from a~homogeneous function
$G(x)$, which has to be singular (not continuous) at the origin, if it is not
constant.

 The lhs of \eqref{rg-defrad} shows immediately that for $F$---a free
Maxwell field, the radial gauge is a~Lorenz gauge. Let $A(x)$ be a Lorenz
gauge as specified in Section \ref{free}. Then one finds
\begin{equation}\label{rg-AVr}
 A^\rg(x)=-\frac{1}{2\pi}\int \dV^\rg(x\cdot l,l)\,\dm^2l\,,
\end{equation}
with
\begin{equation}\label{rg-Vr}
 \dV_b^\rg(s,l)=\dV_b(s,l)+l_b\,\frac{\p\cdot V(s,l)-\p\cdot V(0,l)}{s}\,.
\end{equation}
This is shown by noting that by~\eqref{pre-xA} we have
\begin{equation}
 x\cdot A(\la x)=\frac{1}{2\pi\la}\int (\p\cdot V)(\la x\cdot l,l)\,\dm^2l\,,
\end{equation}
so
\begin{equation}
 S^\rg(x)=\frac{1}{2\pi}\int_0^1 \bigg(\int (\p\cdot V)(\la x\cdot l,l)\,\dm^2l\bigg)\frac{\dm\la}{\la}
\end{equation}
and
\begin{align}
 \p_bS^\rg(x)&=\frac{1}{2\pi}\int_0^1\int l_b (\p\cdot \dV)(\la x\cdot l,l)\,\dm^2l\,\dm\la\\[1ex]
 &=\frac{1}{2\pi}\int l_b\frac{(\p\cdot V)(x\cdot l,l)-(\p\cdot V)(0,l)}{x\cdot l}\,\dm^2l
\end{align}
(note that $S^\rg(x)$ is well defined, since $\int(\p\cdot
V)(0,l)\,\dm^2l=0$).

For a classical field, with sufficiently regular $V(s,l)$, the radial gauge
function \eqref{rg-Vr} is regular and the representation \eqref{rg-AVr} is
well defined (e.g., for a smooth $V(s,l)$ both $V^\rg(s,l)$ and $A^\rg(x)$
are smooth), although with marked differences as compared to $A$. In contrast
to $\dV(s,l)$, which satisfies~\eqref{pre-Vbound}, $\dV^\rg(s,l)$ decays only
as $s^{-1}$ for large $|s|$. This has two consequences: first, there is no
null asymptotics of the form \eqref{pre-as} for $A^\rg(x)$; and second, the
estimate \eqref{pre-decA} is replaced by a slower decay
\begin{equation}\label{rg-bA}
 |A^\rg(x)|\leq\con\int\frac{\dm\Omega(l)}{1+|x\cdot l|}\leq\frac{\con}{1+|x^0|+|\x|}\log(2+|\x|)\,,
\end{equation}
which is easily obtained with the use of estimate \eqref{apes-es}.

However, for the quantum version it follows immediately from the
juxtaposition of formulas \eqref{rg-Vr} and \eqref{fqf-comV} that
$V^\rg(s,l)$, and consequently also $A^\rg(x)$, are not properly defined, due
to the appearance of a distributional quantity at a point: $\p\cdot V(0,l)$.
This, in our view, is the fundamental reason for the reported difficulties
and a need for \emph{a posteriori} regularization in attempts to establish
Feynman rules in the radial gauge (see \cite{leu96} and references therein).

\section{Almost radial gauge}\label{arg}

The source of trouble in the radial gauge is the term $\p\cdot V(0,l)$ in the
function $\dV^\rg(s,l)$. Our plan is to modify the definition
\eqref{rg-defrad} so as to eliminate this term. This will only be possible
with some smearing of the central, reference point. The result of this
programme is the construction of an \emph{almost radial gauge} potential
$A_b^\ar(x)$ in formulas \eqref{arg-arg}, \eqref{arg-Var1}, \eqref{arg-Var2}
and \eqref{arg-qfdef} below.

The starting point for the construction is the following variant of the
identity \eqref{rg-ident}
\begin{equation}\label{arg-id}
 (\xi+1)\ax^aF_{ab}\big(x+\xi \ax\big)
 =\frac{\p}{\p\xi}\big[(\xi+1)A_b\big(x+\xi \ax\big)\big]
 -\p^x_b\big(\ax\cdot A\big(x+\xi \ax\big)\big)\,,
\end{equation}
where $\xi$ is a real parameter, and to abbreviate notation we introduced a
usual affine space symbol
\begin{equation}
 \ax=x-a\,.
\end{equation}
To obtain this identity we shift the argument of $A$ and $F$ in
\eqref{rg-ident} by vector~$a$, and write it with $x$ replaced by $y$:
\begin{equation}\label{arg-ident}
 \la y^a F_{ab}(a+\la y)=\frac{\p}{\p\la}\big[\la A_b(a+\la y)\big]-\frac{\p}{\p y^b}(y\cdot A(a+\la y))\,.
\end{equation}
Now the substitutions $y=\ax$ and $\la=\xi+1$ give the result.

\subsection{Potential of fast decay}

We start discussion with the case of $A_b(x)$ a $C^1$-function of fast decay,
together with its derivative. For such function one could think of integrating identity \eqref{arg-id} with respect to $\xi$ on  $[0,\infty)$, which would correspond to the $\la$-integration of \eqref{rg-ident} on $[1,\infty)$. However, for $F$--a free electromagnetic field, which will be our object of study later, this would not be convergent for spacelike $\ax$. In that case, one has to exploit the oddness (in the spacetime) of its spacelike tail (and evenness of the tail of $A$, see \eqref{pre-tail}) and define integrals as appropriate limits. Anticipating this, we multiply \eqref{arg-id} by
$(-\tfrac{1}{2})\sgn(\xi)$ and integrate with respect to $\xi$ on $\mR$, which results in
\begin{equation}\label{arg-arga}
 -\tfrac{1}{2}\int_\mR\sgn(\xi)(\xi+1)\ax^aF_{ab}\big(x+\xi \ax\big)\dm\xi
 =A_b(x)-\p^x_b S(a,x)\,,
\end{equation}
\begin{equation}\label{arg-Sa}
\begin{aligned}
 S(a,x)
 &=-\tfrac{1}{2}\int_\mR\ax\cdot A\big(x+\xi\ax\big)\sgn(\xi)\dm\xi\\
 &=-\tfrac{1}{2}\int_\mR\ax\cdot A\big(a+\la\ax\big)\sgn(\la-1)\dm\la\,.
\end{aligned}
\end{equation}
Using a technique similar to that applied in Appendix \ref{fr} to analyse function $r^a(x,z)$ defined below, it is easy to show that $S(a,x)$ is regular outside $x=a$, bounded by a
constant, while $|\p^xS(a,x)|\leq\con(1+|a|)|x-a|^{-1}$ (for that, use the second formula in \eqref{arg-Sa}). In order to remove the
singularity, we smear \eqref{arg-arga} with a~real Schwartz function $\rho(a)$
on Minkowski space, such that
\begin{equation}\label{arg-rho}
 \int_M\rho(a)\,\dm a=1\,,\quad  \int_M \rho(a)a^\alpha \dm a=0\quad\text{for}\quad 1\leq|\alpha|\leq n\,,
\end{equation}
for some arbitrarily chosen (large) $n\in\mN\cup\{\infty\}$. The smearing of $\p^xS(a,x)$
may be pulled under the differential sign, and we obtain
\begin{align}
 \int_M S(a,x)\rho(a)\,\dm a
 &=\tfrac{1}{2}\int_M A(y)\cdot\yx\int_\mR\rho\big(x+\xi^{-1}\yx\big)|\xi|^{-5}\dm\xi\,\dm y\\
 &=\tfrac{1}{2}\int_M A(y)\cdot\yx\int_\mR \rho(x+u\,\yx)|u|^3\dm u\,\dm y\,.
\end{align}
In the first step above we substituted $S(a,x)$ \eqref{arg-Sa} and changed
integration variable $a$ to $y=x+\xi(x-a)$. In the second step we substituted
$\xi=u^{-1}$.

We now introduce the vector function
\begin{equation}\label{arg-r}
 r^a(x,z)=\tfrac{1}{2}z^a\int_\mR \rho(x+uz)|u|^3\,\dm u\,,
\end{equation}
which together with its derivatives with respect to $x$ is estimated as in
\eqref{fr-est}, and satisfies the distributional equation (see Appendix
\ref{fr})
\begin{equation}\label{arg-divr}
 \p^z\cdot r(x,z)=\delta(z)\,,
\end{equation}
and define the {\bf almost radial gauge} by
\begin{align}
 A^\ar_b(x)
 &=-\tfrac{1}{2}\int_M\rho(a)\int_\mR\sgn(\xi)(\xi+1)\ax^aF_{ab}\big(x+\xi \ax\big)\,\dm\xi\,\dm a\label{arg-ar1}\\[2ex]
 &=A_b(x)-\p_bS^\ar(x)\,,\label{arg-ar2}\\[1ex]
 S^\ar(x)&=\int_M r(x,x-y)\cdot A(y)\,\dm y=\int_M r(x,z)\cdot A(x-z)\,\dm z\,.\label{arg-arS}
\end{align}
Formulas \eqref{arg-divr}, \eqref{arg-ar2} and \eqref{arg-arS} together show
that our almost radial gauge is in the class of generalized `gauge-invariant
gauges' first postulated by Dirac \cite{dir55} in an attempt to obtain
`physical,' out of local potentials (see also \cite{sta00}). The~generalization consists in nontrivial dependence of $r^a(x,x-y)$ on the first
argument. Potential $A^\ar(x)$ is a continuous function, decaying as
$|x|^{-1}$ for large $|x|$, thus
\begin{equation}
 |A^\ar(x)|\leq\frac{\con}{1+|x|}\,.
\end{equation}
To estimate the radial component of $A^\ar$ assume for simplicity that the
support of $\rho(a)$ is contained in a small ball $|a|\leq\delta$, and note
that $\ax^a x^bF_{ab}=\ax^a a^bF_{ab}$ . Then the contraction of
\eqref{arg-ar1} with $x$ gives
\begin{equation}
 |x\cdot A^\ar(x)|\leq\delta\frac{\con}{1+|x|}\,.
\end{equation}
In the present case of a fast decaying function $A$, the `almost radial'
property has only a global aspect, appearance of $\delta$ on the rhs. It is
the case of free field discussed below, where this term has a fuller
justification.

We note that a strict Dirac gauge with $r(0,x-y)$ replacing $r(x,x-y)$ in
\eqref{arg-arS} is also possible, but it does not have the almost radial
property.

\subsection{Free classical field}

We now assume that $F_{ab}(x)$ is a free classical field obtained from the
potential~\eqref{pre-Adl}, where $\dV(s,l)$ satisfies all restrictions
formulated there. The definition~\eqref{arg-arS} may not be directly applied
in this case, but we shall find that an extension of this definition may be
obtained as a limit of a regularized expression. We admit infrared singular
fields, thus $\Delta V(l)$ may be different from zero. For the sake of the
construction of almost radial gauge at the classical level, it will prove
convenient to use the convention $\oV(s,l)$ \eqref{pre-MV}. The choice
\eqref{fqf-Vinout} will be restored at the stage of quantization.

The following auxiliary functions will be needed:
\begin{equation}\label{arg-bg}
\begin{aligned}
 \sigma(s,l)
 &=\tfrac{1}{2}\int_M\sgn(s-a\cdot l)\rho(a)\dm a\,,\\
  \sax(s,l)
  &=\int_M\frac{\rho(a)}{s-a\cdot l}\,\dm a
  =\int_\mR\frac{\dot{\sigma}(\tau,l)}{s-\tau}\,\dm\tau\,,\\
   \vax(s,l)&=\int_M\frac{\rho(a)a}{s-a\cdot l}\,\dm a
   =-\int_\mR\frac{\p\sigma(\tau,l)}{s-\tau}\,\dm\tau
\end{aligned}
\end{equation}
(we assume $\sigma(s,l)$ is defined by the above formula also in a
neighborhood of the light cone, so $\p\sigma(s,l)$ is also unique in this case).
Properties of these functions are discussed in Appendix \ref{bg}. Let
$\chi(x)$ be a~Schwartz function, such that
\begin{equation}
 \chi(-x)=\chi(x)\,,\qquad \chi(0)=1\,,
\end{equation}
and for $\delta>0$ denote (cf.\ \eqref{arg-arS})
\begin{equation}\label{arg-Sfreechi}
 S^\ar_\delta(x)=\int_M \chi(\delta z) r(x,z)\cdot A(x-z)\,\dm z\,,
\end{equation}
with $A(x)$ as described above. Then the following limit exists
\begin{equation}\label{arg-Sfree}
 S^\ar(x)=\lim_{\delta\searrow0}S^\ar_\delta(x)
 =\frac{1}{2\pi}\int\big[\vax(x\cdot l,l)-\sax(x\cdot l,l)x\big]\cdot \oV(x\cdot l,l)\,\dm^2l\,.
\end{equation}
We give the proof of this fact in Appendix \ref{Sar}. Gradient of this
function is easily obtained
\begin{equation}
 \p_bS^\ar(x)
 =\frac{1}{2\pi}\int\Big\{l_b\p_s\big[(\vax(s,l)-\sax(s,l)x)\cdot \oV(s,l)\big]-\sax(s,l)\oV_b(s,l)\Big\}\Big|_{s=x\cdot l}\,\dm^2l\,.
\end{equation}
We use an identity of the form \eqref{pre-xA} to transform
\begin{equation}
 -\frac{1}{2\pi}x^a\int l_b \p_s\big[\oV_a(s,l)\sax(s,l)\big]\big|_{s=x\cdot l}\,\dm^2l
 =\frac{1}{2\pi}\int \p_a\big[l_b \oV^a(s,l)\sax(s,l)\big]\big|_{s=x\cdot l}\,\dm^2l\,.
\end{equation}
Taking into account that $\p_a[l_b\oV^a\sax]=\oV_b\sax+l_b\p\cdot [\oV\sax]$,
we find
\begin{equation}
 \p_bS^\ar(x)
 =\frac{1}{2\pi}\int l_b\Big\{\p_s\big[\vax(s,l)\cdot \oV(s,l)\big]+\p\cdot\big[\sax(s,l)\oV(s,l)\big]\Big\}\Big|_{s=x\cdot l}\,\dm^2l\,.
\end{equation}

We can now extend the definition of the {\bf almost radial gauge} to the free
field:
\begin{equation}\label{arg-arg}
 A^\ar_b(x)=A_b(x)-\p_b S^\ar(x)
 =-\frac{1}{2\pi}\int\dV^\ar_b(x\cdot l,l)\,\dm^2l\,,
\end{equation}
where
\begin{align}
 \dV^\ar(s,l)&=\doV(s,l)+l\,\Big\{\p_s\big[\vax(s,l)\cdot \oV(s,l)\big]
 +\p\cdot \big[\sax(s,l)\oV(s,l)\big]\Big\}\label{arg-Var1}\\[1ex]
 			&= \doV(s,l)+l\,\Big\{\vax(s,l)\cdot\doV(s,l)
 +\sax(s,l)\p\cdot \oV(s,l)\Big\}\,,\label{arg-Var2}
\end{align}
the second form \eqref{arg-Var2} obtained with the use of identity
\eqref{bg-dbg}. Our almost radial gauge is a functional of the smearing
function $\rho$ \eqref{arg-rho}, which serves to define functions $\sax$ and
$\vax$ \eqref{arg-bg}. Note that putting formally $\rho(a)=\delta(a)$ (which,
precisely, is not allowed), one has $\sax(s,l)=s^{-1}$, $\vax(s,l)=0$ and
\eqref{arg-Var2} reproduces formula \eqref{rg-Vr} with the term $(\p\cdot
V)(0,l)$ omitted (which, again, would result in a singularity).

The almost radial gauge is again a Lorenz gauge satisfying the wave equation.
It is also gauge invariant with respect to gauge transformations
\begin{equation}
 A_b(x)\to A_b(x)+\p_b\Lambda(x)\,,\quad \Lambda(x)=-\frac{1}{2\pi}\int\bar{\alpha}(x\cdot l,l)\,\dm^2l\,,
\end{equation}
which correspond to the replacement $\oV_b(s,l)\to
\oV_b(s,l)+l_b\bar{\alpha}(s,l)$. This replacement leaves $\dV_b^\ar(s,l)$
unchanged. To show this we put $l\bar{\alpha}(s,l)$ in place of $\oV(s,l)$ in
\eqref{arg-Var2}. Recall that to calculate $\p\cdot[l\bar{\alpha}(s,l)]$ one
has to extend $l\bar{\alpha}(s,l)$ to the neighborhood of the light cone respecting
orthogonality to $l$, e.g.,\ by
\begin{equation}
 l_a\bar{\alpha}(s,l)\to[l_a\bar{\alpha}(s,l)]_\mathrm{ext}
 =[l_a-(l^0)^{-1}l^2\delta_a^0]\,\bar{\alpha}(s,l)\,,
\end{equation}
and then on the light cone $\p\cdot[l\bar{\alpha}(s,l)]_\mathrm{ext}
=(l\cdot\p+2)\bar{\alpha}(s,l)$. In this way the change in \eqref{arg-Var2}
becomes
\begin{equation}
 l\sax(s,l)\big(s\p_s+l\cdot\p+2\big)\bar{\alpha}(s,l)=0\,,
\end{equation}
where we used identity \eqref{bg-sb} and the homogeneity of degree $-2$ of
$\bar{\alpha}(s,l)$.

Similarly as in the radial gauge case, $\dV^\ar(s,l)$ decays only as $1/s$
for $|s|\to\infty$, so $A^\ar(x)$ has no null asymptotes and decays as
\begin{equation}\label{arg-bA}
 |A^\ar(x)|\leq\frac{\con}{1+|x^0|+|\x|}\log(2+|\x|)\,.
\end{equation}
On the other hand, using identity \eqref{bg-sb} we find
\begin{align}
 &x\cdot \dV^\ar(x\cdot l,l)\\
 =&x\cdot\doV(x\cdot l,l)+(\p\cdot \oV)(x\cdot l,l)
 +\vax(x\cdot l,l)\cdot\big[l\,\p\cdot \oV(x\cdot l,l)+x\cdot l\,\doV(x\cdot l,l)\big]\,,
\end{align}
and by identity \eqref{pre-xA} we obtain
\begin{equation}\label{arg-xA}
 x\cdot A^\ar(x)= -\frac{1}{2\pi}\int \vax(s,l)\cdot\big[l\,\p\cdot \oV(s,l)+s\doV(s,l)\big]\big|_{s=x\cdot l}\,\dm^2l\,.
\end{equation}
The expression in square brackets in \eqref{arg-xA} is bounded by a constant,
while $\vax(s,l)$ satisfies \eqref{bg-dec}, so by the use of estimate \eqref{apes-es} we find
\begin{equation}\label{arg-bxA}
\begin{aligned}
 |x\cdot A^\ar(x)|&\leq\con\int\frac{\dm\Omega(l)}{(1+|x\cdot l|)^n}\\
 &\leq \frac{\con}{1+|x^0|+|\x|}
 \bigg\{\theta(-x^2)+\frac{\theta(x^2)}{(1+|x^0|-|\x|)^{n-1}}\bigg\}\,.
\end{aligned}
\end{equation}
In timelike directions the radial product $x\cdot A^\ar(x)$ vanishes
arbitrarily fast (depending on the choice of the function $\rho(x)$, see
\eqref{arg-rho}). Moreover, let us introduce the scaling
$\rho_\delta(a)=\delta^{-4}\rho(a/\delta)$, which for $\delta\searrow 0$
contracts the function to the central point, and denote by $A^\ar_\delta$ the
corresponding potential. Then \mbox{$\vax_\delta(s,l)=\vax(s/\delta,l)$} and
for $x\cdot A^\ar_\delta(x)$ the bound \eqref{arg-bxA} holds with the
replacement $x\to x/\delta$. Thus outside $x=0$ this quantity vanishes with
$\delta$ as
\begin{equation}
 |x\cdot A^\ar_\delta(x)|\leq\frac{\con\,\delta}{\delta+|x^0|+|\x|}
 \bigg\{\theta(-x^2)+\frac{\theta(x^2)\,\delta^{n-1}}{(\delta+|x^0|-|\x|)^{n-1}}\bigg\}\,.
\end{equation}

\subsection{Smearing with a test function}\label{arg-tf}

Let $K^b(x)$ be a vector Schwartz test function (no conservation condition)
and denote
\begin{equation}\label{arg-WK}
 W(s,l)=\int_M K(x)\delta(s-x\cdot l)\,\dm x\,,
\end{equation}
which is a $C^\infty$-function, fast decaying in $s$, together with all its
derivatives. Then
\begin{equation}\label{arg-AarK}
 A^\ar(K)=\int_M A^\ar(x)\cdot K(x)\,\dm x
 =-\frac{1}{2\pi}\int \dV^\ar(s,l)\cdot W(s,l)\,\dm s\,\dm^2l\,.
\end{equation}
Setting here \eqref{arg-Var1} and integrating by parts (use \eqref{pre-parts}
to transfer the derivative~$\p$), one finds
\begin{equation}\label{arg-AVK}
\begin{aligned}
 A^\ar(K)&=\frac{1}{2\pi}\int \oV(s,l)\cdot \dV_K(s,l)\,\dm s\,\dm^2l\\
 &=-\frac{1}{2\pi}\int \dV(s,l)\cdot \oV_K(s,l)\,\dm s\,\dm^2l
 =\{\oV_K,\oV\}\,,
\end{aligned}
\end{equation}
where
\begin{equation}\label{arg-VK}
\begin{aligned}
 \dV_K(s,l)&=\dW(s,l)+\vax(s,l)l\cdot \dW(s,l)+\sax(s,l)\p(l\cdot W(s,l))\,,\\[1ex]
 &=\dW(s,l)+\p_s[\vax(s,l)l\cdot W(s,l)]+\p[\sax(s,l) l\cdot W(s,l)]\,,\\
 \oV_K(s,l)&=\tfrac{1}{2}\int_\mR\sgn(s-\tau)\dV_K(\tau,l)\,\dm\tau\,.
\end{aligned}
\end{equation}
It follows that
\begin{align}
 l\cdot \oV_K(s,l)&=0\,,\label{arg-VKort}\\
 \Delta V_K(l)&=\p\int_\mR \sax(s,l) l\cdot W(s,l)\,\dm s\,.\label{arg-VKinf}
\end{align}
Orthogonality \eqref{arg-VKort} is showed with the use of identity
\eqref{bg-sb} as follows:
\begin{align}
 l\cdot \dV_K(s,l)&=l\cdot \dW(s,l)+(s\sax(s,l)-1)l\cdot \dW(s,l)
 +\sax(s,l)l\cdot\p(l\cdot W(s,l))\\
 &=\sax(s,l)\big(s\p_s+l\cdot\p\big)(l\cdot W(s,l))=0\,,
\end{align}
the last equality by the homogeneity of $l\cdot W(s,l)$. Equality
\eqref{arg-VKinf} is immediate from the second form of $\dV_K$ in
\eqref{arg-VK}.

We note that the function $\oV_K(s,l)$ satisfies all demands
\eqref{pre-Vbound} with $N=\infty$, \eqref{pre-VdFi} and \eqref{fqf-ortho} on
a test function in the extended symplectic form \eqref{fqf-sympl}, but in
general---for non-conserved $K$---is outside the original test functions
space.

\subsection{More general smearing}\label{arg-mgs}

Now we want to extend the smearing $A^\ar(K)$ to $C^1$-vector functions
$K(x)$, with the spacetime asymptotic behavior characteristic for currents of
scattered free particles or fields. We shall show that with appropriately specific conditions on $K$, the function $W(s,l)$ \eqref{arg-WK} is of
class $C^1$ and satisfies the bounds
\begin{equation}\label{arg-WKb}
 |\dW(s,l)|\leq \frac{\con}{(1+|s|)^{1+\vep}}\,,\quad
 \big|\p [l\cdot W(s,l)]\big|\leq \frac{\con}{(1+|s|)^\vep}\,.
\end{equation}
Therefore, the integrals in \eqref{arg-AVK}, with the definitions \eqref{arg-VK}, are
absolutely convergent for this extension. Moreover, $\oV_K(s,l)$ thus formed
satisfies conditions \eqref{pre-Vbound} with $N=0$, \eqref{pre-VdFi} and
\eqref{fqf-ortho}. In addition, let $g(x)$ be a Schwartz function, with
$g(0)=1$ and denote $K_\delta(x)=g(\delta x)K(x)$. Then we shall find that
irrespective of the shape of $g$ the following extension of the identity
\eqref{arg-AVK} holds true
\begin{equation}\label{arg-limAK}
\begin{aligned}
 A^\ar(K)\equiv\lim_{\delta\searrow0}A^\ar(K_\delta)
 &=\frac{1}{2\pi}\int \oV(s,l)\cdot \dV_K(s,l)\,\dm s\,\dm^2l\\
 &=-\frac{1}{2\pi}\int \dV(s,l)\cdot \oV_K(s,l)\,\dm s\,\dm^2l
 =\{\oV_K,\oV\}\,.
\end{aligned}
\end{equation}

We first note that once the existence of $W(s,l)$ with the properties
\eqref{arg-WKb} is established, the existence of $\oV_K(s,l)$ with the
property \eqref{pre-Vbound} with $N=0$, and orthogonality \eqref{arg-VKort}
are obviously satisfied.

To specify the class for which the above results hold, it is convenient to
characterize such class by splitting the function $K$ into three
$C^1$-con\-tri\-bu\-tions: \mbox{$K=K_1+K_2+K_3$}, with the following
properties:
\begin{gather}
 \begin{gathered}\label{arg-K1}
 K_1(x)=x\, \kappa(x)\,,\quad \supp\kappa\subseteq\{x^2\geq0,\ |x^0|\geq \tfrac{1}{2}\}\,,\\[1ex]
 \kappa(\la x)=\la^{-4}\kappa(x)\qquad \text{for}\qquad \la\geq1\,,\ |x^0|\geq1\,,\\
 |\kappa(x)|\leq\frac{\con}{(1+|x|)^4}\,,\quad
 |\p\kappa(x)|\leq\frac{\con}{(1+|x|)^5}\,,
 \end{gathered}\\
 \supp K_2\subseteq\{x^2\geq0\}\,,\quad |K_2(x)|\leq\frac{\con}{(1+|x|)^3}\,,\quad
 \text{oscillatory part}\,,\label{arg-K2}\\
 |K_3(x)|\leq\frac{\con}{(1+|x|)^{3+\vep}}\,,\quad
 |\p K_3(x)|\leq\frac{\con}{(1+|x|)^{4+\vep}}\,.\label{arg-K3}
\end{gather}
Condition \eqref{arg-K1} characterizes the dominant asymptotic behavior of
particles, and the non-oscillatory part of asymptotic behavior of fields. The
oscillatory contribution in the case of fields is represented by $K_2$, bounded by \eqref{arg-K2}, and we shall characterize it more precisely below. The rest is represented by $K_3$, conditions \eqref{arg-K3} characterizing the next to leading behaviour inside the light cone, and decay of $K$ outside.\footnote{Suppose, for definiteness, that $K$ asymptotically behaves as the current of a free classical Dirac field. Then by standard stationary phase methods (e.g. \cite{rs79}) the rest, beside the leading order in $K$, represented by $K_3$, is rather safely estimated by \eqref{arg-K3}. The leading order in the asymptotic field, by this method, has the form of a homogeneous function of degree $-3/2$, with the support inside the light cone, multiplied by an oscillating term. In the hermitian form of the current, the oscillations cancel for terms with same energy sign, and add up for terms with opposite energy sign. Proportionality of the homogeneous term to $x$ is related to the fact, that the current asymptotically is interpreted as particles escaping in timelike directions (see e.g.\cite{her95}). Matrix elements of the current will have similar structure in the quantum case.}

We prove our claims separately for each contribution $K_i$, except for property
\eqref{pre-VdFi}, which will be discussed at the end.

We first note that all $W_i(s,l)$, defined by \eqref{arg-WK} with $K$
replaced by $K_i$, are absolutely convergent and bounded. Moreover, in case
$i=3$ one has
\begin{equation}\label{arg-WK3}
 |W_3(s,l)|\,,\ |L_{ab} W_3(s,l)|\leq \frac{\con}{(1+|s|)^\vep}\,,\quad
 |\dW_3(s,l)|\leq \frac{\con}{(1+|s|)^{1+\vep}}\,;
\end{equation}
all these statements are proved as in Lemma 20 and Theorem 21 in ref.\
\cite{her21}. Therefore, bounds \eqref{arg-WKb} are satisfied for $W_3$.
Next, we observe that both
\begin{equation}\label{arg-AK3}
 A^\ar(x)\cdot K_3(x)\quad \text{and}\quad A^\ar(x)\cdot K_1(x)=x\cdot A^\ar(x)\,\kappa(x)
\end{equation}
are absolutely integrable (see \eqref{arg-bA} and \eqref{arg-bxA}), so the
$\delta$-regularization in \eqref{arg-limAK} is not needed for these
contributions. Thus, in case $i=3$ we obtain
\begin{align}
 A(K_3)&=-\frac{1}{2\pi}\int_M\int \dV^\ar(x\cdot l,l)\,\dm^2l\, K_3(x)\,\dm x\\
 &=-\frac{1}{2\pi}\int \dV^\ar(s,l)\cdot W_3(s,l)\,\dm s\,\dm^2l\,,
\end{align}
so the thesis for this contribution follows.

In case $i=1$, we denote
\begin{equation}
 U(s,l)=\int_M \kappa(x)\delta(s-x\cdot l)\,\dm x\,,\qquad |U(s,l)|\leq\frac{\con}{1+|s|}
\end{equation}
(the bound again by Lemma 20 in \cite{her21}), and observe that
\begin{equation}\label{arg-UW}
 \p U(s,l)=-\dW_1(s,l)\,,\quad sU(s,l)=l\cdot W_1(s,l)\,.
\end{equation}
Moreover, we note that the interior of the light cone (past and future) may be
parametrized by $x=\la v$, $\la\in\mR$ and $v$ on the future hyperboloid
$v^2=1$, $v^0\geq0$, and then \mbox{$\dm x=|\la|^3\dm\la\,\dm\mu(v)$}, where
$\dm\mu(v)=\dm^3v/v^0$. It follows then from assumptions~\eqref{arg-K1} that
for $|s|\geq l^0$ one has
\begin{equation}\label{arg-W1as}
\begin{aligned}
 W_1(s,l)&=\int \sgn(\la)\,v\,\kappa\big(\sgn(\la)v\big)\delta(s-\la v\cdot l)\,\dm\la\,\dm\mu(v)\\
 &=\sgn(s)\int\kappa\big(\sgn(s)v\big)\frac{v}{v\cdot l}\,\dm\mu(v)\,,
\end{aligned}
\end{equation}
\begin{equation}\label{arg-lW1as}
 l\cdot W_1(s,l)=\sgn(s)\int\kappa\big(\sgn(s)v\big)\,\dm\mu(v)\,.
\end{equation}
Therefore,
\begin{equation}
 \dW_1(s,l)=0\,,\quad \p[l\cdot W_1(s,l)]=0\quad \text{for}\quad |s|\geq l^0\,.
\end{equation}
We can now use \eqref{arg-xA} to obtain (we omit the arguments $(s,l)$ for
the sake of clarity)
\begin{equation}\label{arg-AK1}
\begin{aligned}
 A(K_1)&=-\frac{1}{2\pi}\int \vax\cdot\big[l\,\p\cdot \oV+s\doV\big]U\,\dm s\,\dm^2l\\
 &=-\frac{1}{2\pi}\int \big[sU\p_s(\oV\cdot \vax)+(1-s\sax)\oV\cdot \p U\big]\,\dm s\,\dm^2l\,,
 \end{aligned}
\end{equation}
where for the second equality we integrated $\p$ by parts and used the
equality $\p(l\cdot \vax)=-s\dot{\sax}$, which follows from identities
\eqref{bg-dbg} and \eqref{bg-sb}. Performing now the substitutions
\eqref{arg-UW} (for $sU$ using the second of these relations in both places
where it appears) and integrating $\p_s$ by parts, we obtain
\begin{equation}
 A(K_1)=\frac{1}{2\pi}\int \oV\cdot\big[\dW_1 +\vax l\cdot\dW_1+\sax \p(l\cdot W_1)\big]\,\dm s\,\dm^2l\,,
\end{equation}
which is the required result.

We turn to the case $i=2$. The decay of $K_2$ is not sufficient to apply the
proof presented in case $i=3$. However, the oscillatory behavior of $K_2$
damps the integrals, and we now add further assumption that $W_2$ satisfies
bounds as in case $W_3$:
\begin{equation}\label{arg-WK2}
 |W_2(s,l)|\,,\ |L_{ab} W_2(s,l)|\leq \frac{\con}{(1+|s|)^\vep}\,,\quad
 |\dW_2(s,l)|\leq \frac{\con}{(1+|s|)^{1+\vep}}\,.
\end{equation}
We denote
\begin{equation}\label{arg-W2deldef}
 W_{2\delta}(s,l)=\int_M g(\delta x)K_2(x)\delta(s-x\cdot l)\,\dm x\,,
 \end{equation}
and assume further that
\begin{equation}\label{arg-W2delta}
 |W_{2\delta}(s,l)|\leq \frac{\con}{(1+|s|)^\vep}\,.
\end{equation}
With these assumptions we have
\begin{align}
 A(K_{2\delta})&=\int \dV^\ar(s,l)\cdot W_{2\delta}(s,l)\,\dm s\,\dm^2l\\
 &\to \int \dV^\ar(s,l)\cdot W_2(s,l)\,\dm s\,\dm^2l\quad\text{for}\quad \delta\searrow0\,,
\end{align}
the limit by the dominated convergence theorem. This, together with the
estimates \eqref{arg-WK2}, leads to the thesis. We show in Appendix \ref{osc}
that our assumptions on $W_2$ are satisfied for a term of the type
characteristic for the Dirac field.

Finally, we shall close the proof of our claims by showing \eqref{pre-VdFi}, namely
\begin{equation}\label{arg-VFi}
 \Delta V_K(l)=-\p\Phi_K(l)\,,
\end{equation}
with
\begin{equation}\label{arg-Fi}
\begin{aligned}
 \Phi_K(l)
 &= \int [\kappa(v)+\kappa(-v)]\rho(a)\log\frac{|a\cdot l|}{v\cdot l}\,\dm\mu(v)\,\dm a\\
 &-\int \sax(s,l)l\cdot \big[W(s,l)-W(\sgn(s)\infty,l)\big]\,\dm s\,.
\end{aligned}
\end{equation}
First, we observe that it is only $W_1$ that contributes to $W(\pm\infty,l)$,
so by \eqref{arg-W1as} we have
\begin{equation}
 W(\pm\infty,l)=\pm\int \kappa(\pm v)\frac{v}{v\cdot l}\,\dm\mu(v)\,,\quad
 l\cdot W(\pm\infty,l)=\pm\int\kappa(\pm v)\,\dm\mu(v)\equiv Q_\pm\,.
\end{equation}
Differentiating the first line in \eqref{arg-Fi} we obtain
\begin{align}
 \int [\kappa(v)+\kappa(-v)]\rho(a)\Big[\frac{a}{a\cdot l}&-\frac{v}{v\cdot l}\Big]\,\dm\mu(v)\,\dm a\\
 &=\vax(0,l)(Q_- -Q_+)-W(+\infty,l)+W(-\infty,l)\,.
\end{align}
Differentiation of the second line in \eqref{arg-Fi} gives
\begin{equation}
 -\int \p[\sax(s,l) l\cdot W(s,l)]\,\dm s+\int \p\sax(s,l)\,Q_{\sgn(s)}\,\dm s\,.
\end{equation}
Using \eqref{bg-dbg} in the second integral and summing the contributions we
obtain
\begin{equation}
 \p\Phi_K(l)=-W(+\infty,l)+W(-\infty,l)-\int  \p[\sax(s,l) l\cdot W(s,l)]\,\dm s\,,
\end{equation}
which substituted into \eqref{arg-VFi} gives the correct value. Note that the
differential $\p$ cannot be extracted outside the integral in the last
formula, as without it the integrand is not integrable.

\subsection{Quantum field}\label{arg-qf}

We have shown in the classical case that $A^\ar(K)=\{\oV_K,\oV\}$. However,
recall the identity \eqref{fqf-VbarV}, which holds in each of the cases
\eqref{fqf-Vinout}, that is both for `in' and for `out' algebra. Thus, on the
algebraic level, in each of these two cases we propose to \emph{define} the
{\bf quantum almost radial gauge} of the potential by\footnote{Our extended field algebra of Section \ref{fqf} is defined directly in terms of test functions $V(s,l)$, and in that sense the definition \eqref{arg-qfdef} is fully compatible with its classical prototype. We do not consider here the question of convergence of the quantum version of the limit~\eqref{arg-limAK}.}
\begin{equation}\label{arg-qfdef}
 A^\ar(K)=\{V_K,V^\qu\}\,,
\end{equation}
whenever $V_K(s,l)$ defined by \eqref{arg-VK} and \eqref{fqf-Vinout} falls in
our extended symplectic space defined after \eqref{fqf-JJcom}.  This
condition demands stronger regularity properties of $K$ than those assumed in
Section \ref{arg-mgs}, leading to classical version of \eqref{arg-qfdef}. One
can show that a class of such functions may be obtained by a modification of
the currents $J$ considered in \cite{her08}, where the modification consists
in demanding the continuity equation to be satisfied only asymptotically in
time (in the form of an appropriately regular analogue of \eqref{arg-K1}). On the other hand, in concrete representations the scope of admitted
test functions is wider, and the definition of the almost radial gauge may be
also appropriately extended at the level of representation. We do not go into
details here and leave the question to be decided in applications.

The almost radial gauge is a Lorenz gauge. For $K(x)=\p F(x)$ one finds that
$V_K(s,l)=W(s,l)\propto \,l$,  which is in the zero equivalence class. Thus
\begin{equation}
 A^\ar(\p F)=0\,.
\end{equation}
Moreover, for $K(x)=\Box G(x)$ we have $W(s,l)=0$, so
\begin{equation}
 A^\ar(\Box G)=0\,.
\end{equation}

For test functions $K(x)$ in the class of conserved currents admitted in
$A(K)$ in one of the cases `in' or `out' one has \mbox{$l\cdot W(s,l)=0$}, so
that  \mbox{$\dV_K(s,l)=\dW(s,l)$} and $V_K(s,l)=W(s,l)$. Therefore, in this
case
\begin{equation}
 A^\ar(K)=A(K)\,,
\end{equation}
which reproduces the extended free algebra.

Finally, let us note the following important corollary. If $K$ is a
conserved, non-radiating current, such as the current of a free, charged
classical particle, or of the free (classical or quantum) Dirac field, then
$W(s,l)=W(l)$ is constant in $s$ and $l\cdot W(l)=Q$, the charge of the
current (see ref.\ \cite{her95}, discussion starting with Eq.\
(2.70)). Therefore, for such currents $V_K=0$ and
\begin{equation}\label{arg-AK0}
 A^\ar(K)=0\,.
\end{equation}

\subsection{Transformation and commutation properties}\label{arg-trancom}

We recall that almost radial gauge is a functional of the function $\rho$
serving to smear the point $a$ from which the radial line to $x$ is drawn.
Let us make this dependence explicit by writing $A^\ar(K;\rho)$. With
Poincar\'e transformations of the extended algebra defined in
\eqref{fqf-poinc}, these elements have the covariance property given by
\begin{equation}\label{arg-alpha}
 \alpha_{z,\Lambda}\big[A^\ar(K;\rho)\big]=A^\ar\big(T_{z,\Lambda}K;T_{z,\Lambda}\rho\big)\,.
\end{equation}
In this sense,  $A(K;\rho)$ may be regarded algebraically covariant, with
unitarily implementable translations. For the proof of the covariance
relation one notes that
\begin{align}
 W[T_{z,\Lambda}K](s,l)&=\Lambda W[K](s-z\cdot l,\Lambda^{-1}l)\,,\\
 \sax[T_{z,\Lambda}\rho](s,l)&=\sax[\rho](s-z\cdot l,\Lambda^{-1}l)\,,\\
 \vax[T_{z,\Lambda}\rho](s,l)
 &=\Lambda\vax[\rho](s-z\cdot l,\Lambda^{-1}l)+z\sax[\rho](s-z\cdot l,\Lambda^{-1}l)\,.
\end{align}
Setting these relations into the formula \eqref{arg-VK} for $V_K$, and then
using the result in \eqref{fqf-poinVV}, one arrives at the thesis.

Elements $A^\ar(K;\rho)$ do not have compact localization even if the support
of $K$ is compact. However, the following remnant timelike locality with
respect to $A$ holds. Let the support of an admissible test vector function
$K$ and the support of a conserved test current $J$ be timelike separated,
and let $V(s,l)$ for current $J$ \eqref{pre-VJ} have $\Delta V(l)=0$ (in
particular, this is always true for compactly supported $J$). Then
\begin{equation}\label{arg-AarA}
 [A^\ar(K;\rho),A(J)]=0\,,
\end{equation}
irrespective of the choice of $\rho$.  To show this, we first note that it is
sufficient to show this when the two supports are placed inside the future
and the past parts of the same light cone, with any vertex point $b$. Next, we
apply to the commutator translation automorphism $\alpha_{-b,\id}$. The
commutator, without changing its value (being proportional to identity),
takes now the form $[A^\ar(K';\rho'),A(J')]$, where $K'$ and $J'$ are
similarly separated as $K$ and $J$, but with the vertex of the light cone in the
origin. In consequence, the functions $W'(s,l)$ for $K'$ \eqref{arg-WK}, and
$V'(s,l)$ for $J'$ \eqref{pre-VJ}, are supported in $s\in(0,+\infty)$ or
$s\in(-\infty,0)$, each in a different of the two sets. Now, it follows from
\eqref{arg-VK} that the support of $\dV_{K'}(s,l)$ is not larger than the
support of $W'(s,l)$, so the same conclusion is valid for the pair
$\dV_{K'}(s,l)$ and $V'(s,l)$. But from $\Delta V'(l)=0$, together with
\eqref{fqf-Vinout}, it follows that
\begin{equation}
 \{V_{K'},V'\}=\frac{1}{2\pi}\int_\mR \dV_{K'}(s,l)\cdot V'(s,l)\,\dm s\,\dm^2l\,,
\end{equation}
so the thesis follows.

\section{Almost radial gauge and string-localized fields: \protect\linebreak comparative remarks}\label{dis}

Definition of the almost radial gauge was motivated by integration, on the
classical level, of the electromagnetic potential along straight axes
\eqref{arg-ar1}. Another construction based on integration along infinitely
extended straight lines is the formalism of string-localized fields
(potentials) developed by Mund, Schroer and Yngvason \cite{msy06} (based on
earlier works, as explained in this reference) and worked out later in many
articles.\footnote{See e.g.\ a recent discussion of an application to QED
problems \cite{mrs20}.} The two constructions result from quite different
motivations and aims; nevertheless, it seems interesting to make some
comparison. For that purpose we briefly recall and formulate in our language
the latter in the case of electrodynamics.\footnote{Our perspective is different
from the discussion of string-localized fields in \cite{msy06}, which starts
from the unitary representations of the Poincar\'e group and engages the
notion of modular localization \cite{bgl02}. Here we want to avoid the
\emph{a priori} assumption on the action of a representation of the Lorentz
group.} All electromagnetic fields are assumed to satisfy the homogeneous
Maxwell equations.

The string-localized potentials in electrodynamics are distributions in
position vectors $x$ and spacelike unit vectors $e$ defined by\footnote{There
exists a version of this construction with lightlike $e$, see \cite{gmv18}.}
\begin{equation}\label{dis-defA}
 A^\slf_b(x,e)=\int_{\mR_+}F_{ba}(x+\la e)e^a\,\dm \la\,.
\end{equation}
To obtain a vector potential as an $x$-distribution, one needs a scalar test
function $h(e)$ on the hyperboloid $e^2=-1$, such that
\begin{equation}\label{dis-h}
 \int h(e)\,\dm\nu(e)=1\,,
\end{equation}
where the integration is over the hyperboloid with the Lorentz invariant
measure $\dm\nu(e)$. Then
\begin{equation}\label{dis-defsmear}
 A^\slf_b(x;h)=\int A^\slf_b(x,e)h(e)\,\dm\nu(e)
\end{equation}
is an $h$-dependent potential, commuting with the field $F_{ab}(y)$ at points
$y$ in spacelike position to the set of points $x+\la e$, $\la\in\mR_+$,
$e\in\supp h$. In particular, if $\supp h$ is contained in some double-cone
spacelike separated from $0$, then $A^\slf(x;h)$ is localized in a spacelike
cone.\pagebreak[2]

This potential is in the Lorenz class, and using the representation
\eqref{pre-F} we show in Appendix \ref{string} that on the classical level
\begin{equation}\label{dis-astring}
 A^\slf_b(x;h)=-\frac{1}{2\pi}\int \dV^\slf_b(x\cdot l,l)\,\dm^2l\,,
\end{equation}
where
\begin{equation}\label{dis-vstring}
 \dV^\slf(s,l)=\dV(s,l)-l\, n(l)\cdot \dV(s,l)
 -l z(l)\cdot \int\frac{\dV(\tau,l)}{s-\tau}\,\dm\tau\,,
\end{equation}
\begin{equation}
 n(l)=\int\frac{eh(e)}{e\cdot l}\,\dm\nu(e)\,,\qquad
 z(l)=\int eh(e)\delta(e\cdot l)\,\dm\nu(e)\,,
\end{equation}
where the integrals in \eqref{dis-vstring} and in $n(l)$ are in the principal value sense,
and not to burden notation we do not indicate the dependence on $h$
explicitly. For a vector Schwartz test function $K(x)$ we then find in
analogy to \eqref{arg-AVK}
\begin{equation}\label{dis-Aslf}
 A^\slf(K;h)=\{V^\slf_K,V\}\,,
\end{equation}
where in place of \eqref{arg-VK}, and keeping notation \eqref{arg-WK}, we
have
\begin{equation}\label{dis-Vslf}
 V^\slf_K(s,l)= W(s,l)-n(l)\,l\cdot W(s,l)+z(l)\int\frac{l\cdot W(\tau,l)}{s-\tau}\,\dm\tau\,.
\end{equation}
In contrast to $W(s,l)$, which is of fast decrease, $V^\slf_K(s,l)$ vanishes
in general only as $s^{-1}$, see the last term in \eqref{dis-Vslf}.
Therefore, the (exponentiated) formula \eqref{dis-Aslf} cannot give, upon
quantization, an element in the local algebra, even if $K$ is local. However,
we have
\begin{equation}
 \wt{V}^\slf_K(\w,l)
 =\wt{W}(\w,l)-n(l)\,l\cdot \wt{W}(\w,l)+i\pi z(l)\sgn(\w)l\cdot\wt{W}(\w,l)\,,
\end{equation}
which is sufficient for elements \eqref{dis-Aslf} to be represented in the
standard vacuum representation of the field $F$, in contrast to the almost
radial gauge, which needs the extended algebra including infrared singular
fields.

The formulation of a gauge theory in terms of string-localized potentials
puts stress on their Poincar\'e covariance. In vacuum representation the
action of the Poincar\'e automorphisms $\alpha_{z,\Lambda}$ of the observable
algebra given by \eqref{fqf-poinc} is implemented by a unitary representation
$U(z,\Lambda)$. Then, in this representation, one also has
\begin{equation}\label{dis-lorsl}
 \alpha_{z,\Lambda}\big[A^\slf(K;h)\big]=A^\slf\big(T_{z,\Lambda}K; T_{0,\Lambda}h\big)
 =U(z,\Lambda)A^\slf(K;h)U(z,\Lambda)^*\,.
\end{equation}
At the algebraic level our transformation law \eqref{arg-alpha} is analogous
to \eqref{dis-lorsl}. However, as mentioned before, in our case not all
Lorentz transformations are implemented.

Let us now return again to the question of localization and commutation.
String-localized fields located on spacelike separated strings commute (which
remains true for slightly smeared strings); this is an important property in
the formalism of these fields. Whether almost radial gauge may be localized
in spacelike extended regions in some special cases and special points $x$ is
a~subtle question, which we do not want to consider here; anyway, this cannot
be a generic property of this gauge. On the other hand, it has a timelike
commutation property \eqref{arg-AarA}. We have seen that this property
depended on the fact that the support of $\dV_K(s,l)$ was not greater than
that of $W(s,l)$. The latter property does not hold for $V_K^\slf(s,l)$
because of the last term in \eqref{dis-Vslf}. Therefore, for string-localized
fields timelike commutativity does not hold, in general.

It is interesting to note that the function $z(l)$ vanishes for even
functions~$h$, $h(-e)=h(e)$, which are admitted by the condition
\eqref{dis-h}. For such functions  the last term in \eqref{dis-Vslf} is
absent and the elements $A^\slf(K;h)$ acquire the timelike locality as
formulated above for $A^\ar(K;\rho)$. As here the algebra is the standard
local electromagnetic algebra of the free field, this leads in fact to the
following implication in the vacuum representation. If the support of the
smearing function $K$ is contained in an open double-cone with vertices $x_1$
and $x_2$, $x_2^0>x_1^0$, then $A^\slf(K)$ belongs to the intersection of the
von Neumann algebras connected with the following two regions: the interior of
the future light cone with the vertex in $x_1$, and the interior of the past
light cone with the vertex in $x_2$. This follows from the timelike duality
for the vacuum representation of massless fields, see \cite{buch78}. Condition of
evenness of $h$ may be equivalently formulated as string integration along
whole real axes, instead of semi-axes; the spacelike localization and
commutation properties are then retained (with obvious
modifications).\footnote{The relation with the notion of modular
localization, as used in the string-localized formalism of \cite{msy06}, is
then lost, or at least becomes indirect.}

One could also observe that the last term in \eqref{dis-Vslf} seems to worsen
the infrared/adiabatic limit behavior: for $K$ becoming infinitely extended so
that $W(s,l)$ does not vanish in infinity, this term becomes meaningless. On
the other hand, if the term is absent, then the algebra of elements
\eqref{dis-Aslf} may be also built in the extended algebra of electromagnetic
field described earlier.

\section{Outlook}\label{outlook}

The results of the analysis in \cite{her95,her21} suggest that, for the classical Maxwell-Dirac system in an appropriate radial-like gauge, the asymptotic Dirac and radiation fields may be largely separated, with the Dirac field carrying its own Coulomb field. Motivated by this, we proposed in \cite{her98, her08} an asymptotic algebra incorporating these ideas on the quantum level. The free electromagnetic part of this algebra is the extended algebra described in Section \ref{fqf}. In order to turn on full interaction in this setting, one needs an appropriate potential. The present construction is a further step in our program.

One should stress that our almost radial gauge could not be constructed within the limits of the standard local theory: see remarks at the end of Section \ref{arg-tf} and the definition of the quantum almost radial gauge in Section~\ref{arg-qf}. Our construction engages nonlocal quantum variables of the extended algebra, which do not appear in the local theory (either in its $C^*$-structure, or in its irreducible representations). Therefore, these nonlocal aspects are not a~consequence, but a prerequisite of the present construction. The long-range structure of the representations of our asymptotic algebra, quite different from the structure in local quantum electrodynamics, was partly considered in \cite{her08, her17}. Further investigation of these questions, not being directly related to the construction of the almost radial gauge, is left for a discussion elsewhere.

\section*{Acknowledgement}

I am grateful to Pawe\l\ Duch, Wojciech Dybalski and Jos\'e M.\
Gracia-Bond\'ia for reading the manuscript and for interesting
correspondence.
Also, some editorial remarks from the Referees are gratefully acknowledged.

\setcounter{subsection}{0}
\renewcommand{\thesubsection}{\Alph{subsection}}

\section*{Appendix}

\subsection{Estimates}\label{apes}

Here we note the following estimates. For $a,b,c,\alpha$, all $>0$, one has
\begin{equation}\label{apes-es}
 \int_0^c\frac{\dm u}{(a+bu)^\alpha}\leq
 \left\{
 \begin{aligned}
 &\frac{\al}{\al-1}\frac{c}{a^{\al-1}(a+bc)}\,,& &\al>1\,,\\
 &\frac{1}{1-\al}\frac{c}{(a+bc)^\al}\,,& &\al<1\,,\\
 &\frac{2c}{a+bc}\log\Big(e+\frac{bc}{a}\Big)\,,& &\al=1\,,
 \end{aligned}\right.
\end{equation}
with Euler's number $e$. Cases $\alpha\lessgtr 1$ may be found in
\cite{her95}, Appendix B, and are reproduced here for convenience of the
reader. In case $\alpha=1$ the result of integration is
$\dsp\frac{1}{b}\log\Big(1+\frac{bc}{a}\Big)$. For $bc\leq a$ this is bounded by
$\dsp\frac{c}{a}\leq\frac{2c}{a+bc}$, while for $bc\geq a$ by
$\dsp\frac{2c}{a+bc}\log\Big(1+\frac{bc}{a}\Big)$.

\subsection{Function $r^a(x,z)$}\label{fr}

Here we discuss properties of function $r^a(x,z)$ defined in \eqref{arg-r},
and we start with its estimate. We denote by $x_z$ and $x_\bot$ the parallel
and the perpendicular (in Euclidean metric) to $z$ components of $x$,
respectively. Then for each $k\in\mN$ we have
\begin{equation}\label{fr-est}
 |r^a(x,z)|\leq \con(k)\frac{(1+|x_z|)^3}{(1+|x_\bot|)^k}\frac{1}{|z|^3}\,.
\end{equation}
To show this, we use $|\rho(x)|\leq \con(1+|x|)^{-k-4}$ to estimate the rhs of
\eqref{arg-r} by
\[
 \con\int_\mR \frac{|z|\,|u|^3\dm u}{(1+|x_\bot|+|u|z|+x_z|)^{k+4}}
 \leq\frac{\con}{|z|^3}\int_0^\infty\frac{(r+|x_z|)^3\,\dm r}{(1+|x_\bot|+r)^{k+4}}\,,
\]
which leads easily to \eqref{fr-est}.

Next, we prove identity \eqref{arg-divr}. We observe that outside $z=0$ the
function \mbox{$\int \rho(x+uz)|u|^3\dm u$} is smooth and homogeneous in $z$
of degree $-4$, so the distribution $\p^z\cdot r(x,z)$ is concentrated in
$z=0$. We choose a $C^1$-function $f(z)$ which together with its derivative
decays at least as $|z|^{-2}$, denote $\hat{z}=z/|z|$ and write $\dm
S(\hat{z})$ for the integration element on the (Euclidean) unit sphere. Then
\begin{align}
 -&\int_M r(x,z)\cdot\p f(z)\,\dm z
 =-\lim_{\ep\searrow0}\int_M r(x,z)\cdot\p f(z)\,\theta(|z|-\ep)\,\dm z\\
 &=\tfrac{1}{2}\lim_{\ep\searrow0}\int_M \int_\mR \rho(x+uz)|u|^3\,\dm u f(z)\delta(|z|-\ep)|z|\,\dm z\\
 &=\tfrac{1}{2}\lim_{\ep\searrow0}\int_{S^3} \int_\mR \rho(x+r\hat{z})|r|^3\,\dm r f(\ep\hat{z})\,\dm S(\hat{z})
 =f(0)\int_M\rho(x+z)\,\dm z=f(0)\,,
\end{align}
where the third equality is obtained by the change of variables $r=\ep u$.
This ends the proof.

\subsection{Functions $\sax$ and $\vax$}\label{bg}

The functions $\sax$ and $\vax$ defined by \eqref{arg-bg} are smooth outside
$l=0$ and homogeneous of degree $-1$ with respect to their arguments, and
satisfy the following identities
\begin{gather}
 \p\sax(s,l)+\dot{\vax}(s,l)=0\,,\label{bg-dbg}\\
 s\sax(s,l)=1+l\cdot\vax(s,l)\,.\label{bg-sb}
\end{gather}
The  Fourier transforms are
\begin{equation}\label{bg-F}
 \wt{\sax}(\w,l)=i\pi\sgn(\w)\hat{\rho}(\w l)\,,\quad
 \wt{\vax}(\w,l)=\pi\sgn(\w)(\p\hat{\rho})(\w l)\,.
\end{equation}
Conditions \eqref{arg-rho} are equivalent to
\begin{equation}\label{bg-rho}
 \hat{\rho}(0)=\frac{1}{2\pi}\,,\quad \p^\alpha\hat{\rho}(0)=0\quad\text{for}\quad 1\leq|\alpha|\leq n\,,
\end{equation}
which implies that $\wt{\vax}(\w,l)$ is $n-1$ times continuously
differentiable and
\begin{equation}
 \p_\w^k\wt{\vax}(\w,l)_{|\w=0}=0\,,\quad k=0,\ldots, n-1\,,
\end{equation}
while $\p_\w^n\wt{\vax}(\w,l)$ is bounded and integrable.  It follows that
\begin{equation}\label{bg-dec}
 |\vax(s,l)|\leq\frac{\con}{(1+|s|)^n}\,.
\end{equation}

\subsection{Derivation of $S^\ar(x)$ for free field}\label{Sar}

Here we prove formula \eqref{arg-Sfree}. Formula \eqref{arg-Sfreechi} for
$S^\ar_\delta(x)$ may be interpreted as the definition \eqref{arg-arS} with
$\chi(\delta z)A(x-z)$ substituted for $A(x-z)$. Equivalently, this is
expression \eqref{arg-Sa} with $\chi(-\delta\xi\ax)A(x+\xi\ax)$ substituted
for $A(x+\xi\ax)$, and smeared with $\rho(a)$. We change integration
variables from $a$ to $z=a-x$, multiply the expression for convenience by
$2\pi$, and obtain
\begin{equation}
 2\pi S^\ar_\delta(x)
 =-\tfrac{1}{2}\int_M\int_\mR \chi(\delta\xi z)\rho(x+z)
 \int z\cdot \dV(x\cdot l-\xi z\cdot l,l)\,\dm^2l\, \sgn(\xi)\,\dm\xi\,\dm z\,.
\end{equation}
We note that the integrand is absolutely integrable, so we can freely change
the order of integration and leave the $\dm^2l$ integral to be performed at
the last stage. In the remaining integrals we change $\xi$ to $\la=\delta\xi$
and reduce integration with respect to $\la$ to the positive semi-axis by
reflecting $\la\to -\la$ from the negative semi-axis. The result may be
written as
\begin{equation}\label{sff-int}
 -\int_M\int_{\mR_+} \chi(\la z)\rho(x+z)
 \frac{\p}{\p\la}\frac{z\cdot Z(x\cdot l,\delta^{-1}\la z\cdot l,l)}{z\cdot l}\,\dm\la\,\dm z\,,
\end{equation}
where
\begin{equation}
 Z(\tau,u,l)=\oV(\tau,l)
 -\tfrac{1}{2}\big[\oV(\tau+u,l)+\oV(\tau-u,l)\big]\,.
\end{equation}
We note that $Z(\tau,u,l)$ is bounded, $Z(\tau,0,l)=0$, and
\begin{equation}
 \lim_{\delta\to0}Z(x\cdot l,\delta^{-1}\la z\cdot l,l)=\oV(x\cdot l,l) \qquad \text{almost everywhere}\,,
\end{equation}
with the choice of $\oV(s,l)$ according to the formula \eqref{pre-MV}.

We now submit \eqref{sff-int} to the following transformations:
\begin{itemize}
\item[--] integration with respect to $\la$ by parts;
\item[--] replacement of $\rho(x+z)$ by
    $\nu(x,z)=\frac{1}{2}[\rho(x+z)+\rho(x-z)]$, possible due to the
evenness in $z$ of the combined remaining factors;
\item[--] choice of variables $w^a$ for vector $z$, such that $w^0=z\cdot
    l$, \mbox{$\dm z=\dm w^0 \dm^3 w_\bot$}, with $w_\bot=(w^1,w^2,w^3)$.
\end{itemize}
In this way \eqref{sff-int} takes the form
\begin{equation}\label{sff-fZ}
 \int_\mR\int_{\mR_+}(w^0)^{-1}f(x,\la,w^0)\cdot Z(x\cdot l,\delta^{-1}\la w^0,l)\,\dm\la\,\dm w^0\,,
\end{equation}
where
\begin{equation}
 f(x,\la,w^0)=\int_{\mR^3} \frac{\p}{\p\la}\chi(\la w)\nu(x,w) w\,\dm w_\bot\,.
\end{equation}
It is easy to see that $f(x,\la,w^0)$ is a $C^\infty$-function, and
$f(x,\la,0)=0$ due to the oddness of the integrand in $w$, so
$(w^0)^{-1}f(x,\la,w^0)$ is also $C^\infty$. An easy calculation also shows
that the latter function is absolutely integrable with respect to
$\dm\la\,\dm w^0$. Therefore, we can take the $\delta\to0$ limit under all
integrations and the limit of \eqref{sff-fZ} becomes
\begin{equation}
 \oV(x\cdot l,l)\cdot\int_\mR\int_{\mR_+} f(x,\la,w^0)(w^0)^{-1}\,\dm\la\,\dm w^0
 =-\oV(x\cdot l,l)\cdot \int_{\mR^4} \frac{w\,\nu(x,w)}{w^0}\,\dm w\,,
\end{equation}
in the latter form integral in the principal value sense. We can now restore
back the abstract notation of vector $z$, and the function $\rho(x+z)$ in
place of $\nu(x,z)$ (due to the evenness of $(w^0)^{-1}w$). We finally
restore the integration variable $a=x+z$ and then
\begin{equation}
 -\int_{\mR^4} \frac{\nu(x,w)w}{w^0}\,\dm w=\int_M\frac{\rho(a)(a-x)}{x\cdot l-a\cdot l}\,\dm a
 =\vax(x\cdot l,l)-x\sax(x\cdot l,l)\,.
\end{equation}

\subsection{Oscillatory asymptotic term}\label{osc}

Here we make more specific assumptions on oscillating smearing function $K_2$
introduced in Section \ref{arg-mgs}. Function $K_2$ specified below obviously
satisfies \eqref{arg-K2}, and we shall show that assumptions \eqref{arg-WK2}
and \eqref{arg-W2delta} are satisfied with $\vep=1$.

Let $K_2(x)$ be of the form (cf.\ \cite{her95})
\begin{equation}
 K_2(x)= e^{i\mu\sqrt{x^2}}k(x)\,,\qquad \mu>0\,,
\end{equation}
where $k(x)$ is a $C^2$-vector function, supported inside the light cone and
vanishing for $|x^0|\leq 1$. Moreover, we assume that in the parametrization
$x=\la v$ introduced after \eqref{arg-UW} the following bounds are satisfied
\begin{equation}\label{osc-bounds}
 |(\p^\alpha k)(\la v)|\leq\frac{\con}{\la^{3+|\alpha|}(v^0)^n}\,,\qquad
 |\alpha|\leq2\,.
\end{equation}
where for simplicity we assume that $n$ may be arbitrarily large; by a rather
tedious control of the decay in $v$ this could be substantially weakened, but
we do not need this for our illustrative purposes.

For $W_{2\delta}$ formed as in \eqref{arg-W2deldef} we have
\begin{equation}
 W_{2\delta}(s,l)=\int e^{i\mu\la}k(\la v)g(\delta\la v)\delta(s-\la v\cdot l)|\la|^3\dm\la\,\dm\mu(v)\,.
\end{equation}
We integrate with respect to $\la$, which gives
\begin{equation}\label{osc-W2del}
 W_{2\delta}(s,l)=\int e^{i\mu \la}k(\la v)g(\delta\la v)|\la|^3\frac{\dm\mu(v)}{v\cdot l}\,,
\end{equation}
where from now on we put $\la=\la(s,v,l)=s/v\cdot l$. It is clear that
$|W_{2\delta}(s,l)|$ is bounded by a constant independent of $\delta$, and
\begin{equation}\label{osc-W2delW2}
 \lim_{\delta\to 0}W_{2\delta}(s,l)=W_2(s,l)
 =\int e^{i\mu \la}k(\la v)|\la|^3\frac{\dm\mu(v)}{v\cdot l}\,.
\end{equation}

To proceed, we need an integral identity on the hyperboloid $v^2=1$, $v^0>0$.
Consider operators
\begin{equation}
 M_{ab}=v_a\frac{\p}{\p v^b}-v_b\frac{\p}{\p v^a}\,,\quad
 \delta_b=v^aM_{ab}=\frac{\p}{\p v^b}-v_bv\cdot\frac{\p}{\p v}\,,
\end{equation}
intrinsic differentiation operators in the hyperboloid. The measure
$\dm\mu(v)$ is Lorentz invariant, and operators $M_{ab}$ generate Lorentz
transformations in the hyperboloid, so if $\chi(v)$ and $M_{ab}\chi(v)$ are
absolutely integrable, then
\begin{equation}
 \int M_{ab}\chi(v)\,\dm\mu(v)=0\,.
\end{equation}
Moreover, we have $M_{ab}v^a\chi(v)=(\delta_b-3v_b)\chi(v)$. Therefore, if
\begin{equation}
 |\chi(v)|\leq\frac{\con}{(v^0)^{3+\vep}}\,,\qquad |\delta\chi(v)|\leq\frac{\con}{(v^0)^{2+\vep}}\,,
\end{equation}
then
\begin{equation}\label{osc-ident}
 \int (\delta-3v)\chi(v)\,\dm\mu(v)=0\,.
\end{equation}

We note the identities and a bound
\begin{equation}
 l\cdot \delta \frac{1}{v\cdot l}=1\,, \quad
 e^{i\mu\la}=-\frac{i}{s\mu}\,l\cdot\delta e^{i\mu\la}\,,\quad
 l\cdot \delta\frac{v}{v\cdot l}=\frac{l}{v\cdot l}\,,\quad
 \frac{1}{v\cdot l}\leq \frac{2v^0}{l^0}\,.
\end{equation}
Substituting the second of these relations in \eqref{osc-W2del} and
integrating by parts by \eqref{osc-ident} (our assumptions on decay of $k$
and $\p k$ make it possible), we find
\begin{align}
 W_{2\delta}(s,l)
 &=\frac{i}{\mu s}\int e^{i\mu \la}g(\delta \la v)|\la|^3\Big(k(\la v)+\frac{\la}{v\cdot l}l\cdot \p k(\la v)\Big)\,\dm\mu(v)\,,\\
 &+\frac{i\delta}{\mu s}\int e^{i\mu \la}|\la|^3 k(\la v)  \frac{\la}{v\cdot l} l\cdot\p g(\delta\la v)\,\dm\mu(v)\,.
\end{align}
We note that $|g(\delta\la v)|$ is bounded by a constant and tends to $1$ for
$\delta\to0$, while $\delta|\la||v^0||\p g(\delta\la v)|$ is also bounded by
a constant and tends to zero in that limit. Thus using \eqref{osc-bounds} we
find that $|W_{2\delta}(s,l)|\leq\con|s|^{-1}$, so the bound
\eqref{arg-W2delta} and the first of the bounds \eqref{arg-WK2} are satisfied
with $\vep=1$. Moreover, taking the limit we find
\begin{equation}
 W_2(s,l)
 =\frac{i}{\mu s}\int e^{i\mu \la}|\la|^3\Big(k(\la v)+\frac{|\la|}{v\cdot l}l\cdot \p k(\la v)\Big)\,\dm\mu(v)\,.
\end{equation}
Differentiating this formula with respect to $l^a$ and $s$ and using similar
technique as above one proves the remaining bounds in \eqref{arg-WK2} for
$|s|\geq1$. For $|s|\leq2$ formula \eqref{osc-W2delW2} should be used
instead. We omit the details.

\subsection{String-localized potential}\label{string}

We prove here formula \eqref{dis-astring} with \eqref{dis-vstring}.
First, we note the following simple limiting formula.
Let $f$ and $g$ be $C^0$-functions on $\mR$, $C^1$ in a neighborhood of $0$,  such that $f$ is bounded, and $g(u)/u$ is absolutely integrable on $|u|\geq1$. Then
\begin{equation}\label{string-fg}
 \lim_{\Lambda\to\infty}\int_\mR\frac{f(\tau)g(\Lambda\tau)}{\tau}\,\dm\tau
 =\lim_{\vep\searrow0}\int_\mR\frac{f(\vep u)g(u)}{u}\,\dm u
 =f(0)\int_\mR\frac{g(u)}{u}\,\dm u\,,
\end{equation}
all integrals in the principal value sense. The first equality is by the change of integration variable $\tau=\vep u$, $\vep\Lambda=1$. The second formula is obviously true for integration restricted to $|u|\geq1$. For $|u|\leq1$ the principal value integral in the middle may be written as
\[
 f(0)\int_{-1}^1\frac{g(u)-g(0)}{u}\,\dm u
 +\vep\int_{-1}^1 \frac{f(\vep u)-f(0)}{\vep u}g(u)\,\dm u\,.
\]
The first term gives the value on the rhs of \eqref{string-fg}, and the second term vanishes in the limit.

Next, we note that formula \eqref{dis-defsmear} may be written as
\[
 A^\slf_b(x,h)=A_b(x)
 +\int\bigg\{\int_{\mR_+}e^a \p_b A_a(x+\la e)h(e)\,
 \dm\la\bigg\}\,\dm\nu(e)\,,
\]
where for simplicity let $h(e)$ be compactly supported. For $e$ in a compact subset of the hyperboloid one has
 $|\p_bA_a(x+\la e)|\leq\con/(1+\la)^2$ (the constant depending on $x$ and the
support in $e$, use e.g. Theorem 18 in \cite{her21}), so the integrand of the iterated integral above is absolutely integrable, and we can change the order of integration. Let us denote
\begin{equation}\label{string-H}
 k^a(\tau,l)=\int e^ah(e)\delta(\tau-e\cdot l)\,\dm\nu(e)\,,
\end{equation}
which is a smooth function of compact support in $\tau/l^0$. Then, using \eqref{pre-Adl} we find
\[
 \int e^a \p_bA_a(x+\la e)h(e)\,\dm\nu(e)
 =-\frac{1}{2\pi}\int l_b\, k(\tau,l)\cdot
 \ddot{V}(x\cdot l+\la\tau,l)\,\dm\tau\,\dm^2 l\,.
\]
Integrating this with respect to $\la$ on $[0,\Lambda]$, we get
\[
 -\frac{1}{2\pi}\int l_b\, k(\tau,l)\cdot
 \frac{\dV(x\cdot l+\Lambda\tau,l)-\dV(x\cdot l,l)}{\tau}\,\dm\tau\,\dm^2l\,.
\]
This may be separated into the difference of two integrals, with $1/\tau$ treated in the principal value sense. Taking the limit $\Lambda\to\infty$ with the use of formula \eqref{string-fg}, setting \eqref{string-H} and integrating with respect to $\tau$, we obtain
\[
 \frac{1}{2\pi}\int l_b\,\bigg\{\int e^ah(e)\delta(e\cdot l)\dm\nu(e)
 \int_\mR\frac{\dV_a(\tau,l)}{x\cdot l-\tau}\,\dm\tau
 +\int\frac{e^ah(e)}{e\cdot l}\,\dm\nu(e)\,\dV_a(x\cdot l,l)\bigg\}\,\dm^2l\,,
\]
which reproduces the gauge terms in \eqref{dis-astring}--\eqref{dis-vstring}.

\end{document}